\documentclass[sigconf]{acmart}
\usepackage{tikz}
\usepackage{amsmath,amsfonts,bm}
\usepackage{mathtools}
\usepackage{comment}
\usepackage[font=bf]{caption}
\usepackage{graphicx}
\usepackage{booktabs}
\usepackage{array,graphicx,multirow}
\usepackage{dblfloatfix}
\usepackage{soul}
\usepackage{hyperref}
\usepackage{hhline}
\usepackage{algorithm}
\usepackage{algorithmic}
\usepackage{enumitem}
\usepackage{diagbox}
\def\smallerspacecaption{\vspace{-2mm}}
\usepackage{cancel}
\theoremstyle{plain}

\theoremstyle{definition}

\newcommand{\mytool}{\textsc{Attrition}}
\usepackage{subfig}
\usepackage[normalem]{ulem}
\def\removesen{1}
\if 1\removesen
\newcommand\autost[1]{\textcolor{red!40!white}{\sout{#1}}}
\else
\newcommand\autost[1]{}
\fi

\usepackage{pifont}
\newcommand{\cmark}{\ding{51}}

\usepackage{lipsum}
\usepackage{dirtytalk}
\definecolor{cadmiumgreen}{rgb}{0.0, 0.42, 0.24}

\usepackage[framemethod=default]{mdframed}
\newmdenv[linecolor=black,backgroundcolor=gray!20]{myframe}
\usepackage[skins]{tcolorbox}
\newtcolorbox{mytcolorbox}[2][width=0.975\columnwidth,halign=flush center,arc is angular]{colback=gray!10,colframe=black,fonttitle=\bfseries,coltitle=black,colbacktitle=white,enhanced,attach boxed title to top center={yshift=-2mm},title={#2},#1}

\newcommand*\cc[1]{\tikz[baseline=(char.base)]{
\textcolor{black}{      \node[shape=circle,draw,inner sep=1pt] (char) {#1};}}}

\usepackage{fancyhdr,blindtext}
\setcopyright{none}
\acmYear{2022}

\usepackage{flushend}

\usepackage{fancyhdr}

\begin{document}

\title[{\sc Attrition}: \underline{A}ttacking ``S\underline{t}atic'' Hardware \underline{Tr}ojan Detect\underline{i}on \underline{T}echniques Us\underline{i}ng Reinf\underline{o}rcement Lear\underline{n}ing]{{\sc Attrition}: \underline{A}ttacking S\underline{t}atic Hardware \underline{Tr}ojan Detect\underline{i}on \underline{T}echniques Us\underline{i}ng Reinf\underline{o}rcement Lear\underline{n}ing}

\author{Vasudev~Gohil, Hao~Guo, Satwik~Patnaik, and Jeyavijayan~(JV)~Rajendran}
\affiliation{\institution{Electrical \& Computer Engineering, Texas A\&M University, College Station, Texas, USA}}
\email{{gohil.vasudev, guohao2019, satwik.patnaik, jv.rajendran} @tamu.edu}

\begin{abstract}
Stealthy hardware Trojans (HTs) inserted during the fabrication of integrated circuits can bypass the security of critical infrastructures.
Although researchers have proposed many techniques to detect HTs, several critical limitations exist, including: (i)~a low success rate of HT detection, (ii)~high algorithmic complexity, and (iii)~a large number of test patterns.
Furthermore, as we show in this work the most pertinent drawback of prior (including state-of-the-art) detection techniques stems from an incorrect evaluation methodology, \textit{i.e.,} they assume that an adversary inserts HTs randomly.
Such inappropriate adversarial assumptions enable detection techniques to claim high HT detection accuracy, leading to a \textit{``false sense of security.''}
\textit{To the best of our knowledge, despite more than a decade of research on detecting HTs inserted during fabrication, there have been no concerted efforts to perform a systematic evaluation of HT detection techniques.}

In this paper, we play the role of a realistic adversary and question the efficacy of HT detection techniques by developing an automated, scalable, and practical attack framework,~\mytool,~using reinforcement learning (RL).
\mytool~evades eight detection techniques (published in premier security venues, well-cited in academia, etc.) across two HT detection categories, showcasing its agnostic behavior.
\mytool~achieves average attack success rates of $47\times$ and $211\times$ compared to randomly inserted HTs against state-of-the-art logic testing and side channel techniques.
To demonstrate~\mytool's ability in evading detection techniques, we evaluate different designs ranging from the widely-used academic suites (ISCAS-85, ISCAS-89) to larger designs such as the open-source \texttt{MIPS} and \texttt{mor1kx} processors to \texttt{AES} and a \texttt{GPS} module. 
Additionally, we showcase the impact of~\mytool~generated HTs through two case studies (privilege escalation and kill switch) on \texttt{mor1kx} processor. 
We envision that our work, along with our released HT benchmarks and models (post peer-review), fosters the development of better HT detection techniques.
\end{abstract}



\maketitle
\thispagestyle{plain}
\pagestyle{plain}

\renewcommand{\headrulewidth}{0.0pt}
\thispagestyle{fancy}
\lhead{}
\rhead{}
\chead{To Appear in 2022 ACM SIGSAC Conference on Computer and Communications Security (CCS), November 2022}
\cfoot{}

\section{Introduction}
\label{sec:introduction}

\subsection{Globalized IC Supply Chain and Threats}
\label{sec:globalized_supply_chain}

Integrated circuits (ICs) are the backbone of modern computing systems. 
Enabling high-performance and low-power ICs requires access to smaller and faster transistors (building blocks of ICs).
The push for continual miniaturization of transistors necessitates reliance on state-of-the-art fabrication facilities, also known as foundries. 
However, commissioning a state-of-the-art foundry incurs astronomical costs~\cite{tsmc3nm,2nmcost}. 
For instance, TSMC, the world's largest contract IC manufacturer, has allocated $\$$40 billion to increase chip production for the state-of-the-art $2$nm technology node~\cite{2nmcost}.
To reduce the design cost and ameliorate marketing constraints, leading IC design companies like Apple, Qualcomm, and NVIDIA operate in a fabless model~\cite{fabless} and outsource IC manufacturing to off-shore, third-party foundries, which could be potentially \textit{untrustworthy} (refer to Appendix~\ref{app:appendix_supply_chain} for more details).
The U.S. \textit{Department of Defense (DoD)} 2022 action plan for securing defense-critical supply chains indicates that 88\% of microelectronic fabrication is performed overseas, representing a substantive security threat~\cite{DoD_supply_chain}.
Such a distributed supply chain has given rise to numerous security concerns ranging from IP piracy~\cite{alkabani2007active,imeson2013securing,yasin2017provably} to the insertion of malicious logic known as hardware Trojans (HTs)~\cite{agrawal2007trojan,rostami2014primer,hicks2010overcoming,waksman2013fanci,yang2016a2,trippel2021bomberman}.

\subsection{Disruptive Impact of Hardware Trojans}
\label{sec:disruptive_impact}

HTs, once inserted during fabrication, cannot be removed. 
Unlike buggy software, where a software patch can help mitigate the ill effects, the damage incurred by a stealthy HT has far-reaching consequences~\cite{hicks2010overcoming,skorobogatov2012breakthrough,becker2013stealthy,yang2016a2}.
For instance, an HT enables privilege escalation~\cite{sturton2011defeating,yang2016a2} or an HT implanted in critical infrastructure (e.g., radar systems) leads to national security threats~\cite{adee2008hunt}.
To highlight the practicality and importance, researchers have discovered and demonstrated HTs in silicon~\cite{skorobogatov2012breakthrough,becker2013stealthy,yang2016a2}.
For example,~\cite{skorobogatov2012breakthrough} found a ``backdoor'' in a military-grade chip, ~\cite{becker2013stealthy} crafted HTs that compromised cryptographically-secure random number generators used in Intel's \textit{Ivy Bridge} processors, and~\cite{yang2016a2} performed privilege escalation using a capacitor-based HT on fabricated chips.
These examples showcase the pernicious effects of HTs in bypassing the security of a system.
Apart from academic endeavors, the \textit{Defense Advanced Research Projects Agency} (DARPA), a research and development agency of the U.S. DoD, is accelerating HT research through programs like \textit{Safeguards against Hidden Effects and Anomalous Trojans in Hardware} (SHEATH)~\cite{SHEATH} and \textit{Automatic Implementation of Secure Silicon} (AISS)~\cite{aiss}, the latter in collaboration with Synopsys, a leading electronic design automation company.

\begin{table*}[ht]
\centering
\caption{HT detection techniques against~\mytool. \textcolor{cadmiumgreen}{\cmark} indicates that~\mytool~evades detection.}
\label{tab:summary_of_HT_detection_techniques}
\resizebox{\textwidth}{!}{
\begin{tabular}{ccccccccc}
\hline
\textbf{Type} & \multicolumn{4}{c}{\textbf{Logic testing}} & \multicolumn{4}{c}{\textbf{Side channel}} \\ \hline
\textbf{Detection Technique} & \multicolumn{1}{c}{MERO~\cite{chakraborty2009mero}} & \multicolumn{1}{c}{GA+SAT~\cite{GA_SAT}} & \multicolumn{1}{c}{TARMAC~\cite{TARMAC_TCAD}} & TGRL~\cite{pan2021automated} & \multicolumn{1}{c}{MERS~\cite{huang2016mers}} & \multicolumn{1}{c}{MERS-h~\cite{huang2016mers}} & \multicolumn{1}{c}{MERS-s~\cite{huang2016mers}} & MaxSense~\cite{lyu2021maxsense} \\ \hline
\textbf{Venue} & \multicolumn{1}{c}{CHES'09} & \multicolumn{1}{c}{CHES'15} & \multicolumn{1}{c}{TCAD'21} & ASP-DAC'21 & \multicolumn{1}{c}{CCS'16} & \multicolumn{1}{c}{CCS'16} & \multicolumn{1}{c}{CCS'16} & TODAES'21 \\ \hline
\textbf{Algorithm} & \multicolumn{1}{c}{\begin{tabular}[c]{@{}c@{}}Random \\ bit-flipping\end{tabular}} & \multicolumn{1}{c}{\begin{tabular}[c]{@{}c@{}}Genetic\\ algorithm\end{tabular}} & \multicolumn{1}{c}{\begin{tabular}[c]{@{}c@{}}Maximal clique\\ enumeration\end{tabular}} & \begin{tabular}[c]{@{}c@{}}Reinforcement\\ learning\end{tabular} & \multicolumn{1}{c}{\begin{tabular}[c]{@{}c@{}}Random\\ bit-flipping\end{tabular}} & \multicolumn{1}{c}{\begin{tabular}[c]{@{}c@{}}Reordering using\\ Hamming distance\end{tabular}} & \multicolumn{1}{c}{\begin{tabular}[c]{@{}c@{}}Reordering\\ using simulation\end{tabular}} & \begin{tabular}[c]{@{}c@{}}Genetic\\ algorithm\end{tabular}\\ \hline
\textbf{\begin{tabular}[c]{@{}c@{}}Largest Design\\ (\# gates)\end{tabular}} & \multicolumn{1}{c}{\begin{tabular}[c]{@{}c@{}}\texttt{s35932}${}^\dagger$\\(6,500)\end{tabular}} & \multicolumn{1}{c}{\begin{tabular}[c]{@{}c@{}}\texttt{s38417}${}^\dagger$\\(22,179)\end{tabular}} & \multicolumn{1}{c}{\begin{tabular}[c]{@{}c@{}}\texttt{MIPS}\\($\approx$25,000)\end{tabular}} & \begin{tabular}[c]{@{}c@{}}\texttt{s35932}${}^\dagger$\\(12,204)\end{tabular} & \multicolumn{1}{c}{\begin{tabular}[c]{@{}c@{}}\texttt{s35932}${}^\dagger$\\(6,500)\end{tabular}} & \multicolumn{1}{c}{\begin{tabular}[c]{@{}c@{}}\texttt{s35932}${}^\dagger$\\(6,500)\end{tabular}} & \multicolumn{1}{c}{\begin{tabular}[c]{@{}c@{}}\texttt{s35932}${}^\dagger$\\(6,500)\end{tabular}} & \begin{tabular}[c]{@{}c@{}}\texttt{MIPS}\\($\approx$25,000)\end{tabular} \\ \hline
\textbf{\begin{tabular}[c]{@{}c@{}}Evaluation\\ Methodology\end{tabular}} & \multicolumn{1}{c}{\begin{tabular}[c]{@{}c@{}}Randomly-\\ inserted HTs\end{tabular}} & \multicolumn{1}{c}{\begin{tabular}[c]{@{}c@{}}Randomly-\\ inserted HTs\end{tabular}} & \multicolumn{1}{c}{\begin{tabular}[c]{@{}c@{}}Randomly-\\ inserted HTs\end{tabular}} & \begin{tabular}[c]{@{}c@{}}Randomly-\\ inserted HTs\end{tabular} & \multicolumn{1}{c}{\begin{tabular}[c]{@{}c@{}}Randomly-\\ inserted HTs\end{tabular}} & \multicolumn{1}{c}{\begin{tabular}[c]{@{}c@{}}Randomly-\\ inserted HTs\end{tabular}} & \multicolumn{1}{c}{\begin{tabular}[c]{@{}c@{}}Randomly-\\ inserted HTs\end{tabular}} & \begin{tabular}[c]{@{}c@{}}Randomly-\\ inserted HTs\end{tabular}\\ 
\cmidrule(lr){1-5} \cmidrule(lr){6-8} \cmidrule(lr){9-9}
\textbf{\begin{tabular}[c]{@{}c@{}}Claimed defense\\Efficacy \end{tabular}} & \multicolumn{1}{c}{$93.5\%$} & \multicolumn{1}{c}{$79.4\%$} & \multicolumn{1}{c}{$93.1\%$} & $96.1\%$ & \multicolumn{3}{c}{\begin{tabular}[c]{@{}c@{}}`... MERS is effective for any Trojan forms/sizes ..."~\cite{huang2016mers}\end{tabular}} & $96.6\%$
\\ \hline
\textbf{~\mytool~(This Work)} & \multicolumn{1}{c}{\textcolor{cadmiumgreen}{\cmark}} & \multicolumn{1}{c}{\textcolor{cadmiumgreen}{\cmark}} & \multicolumn{1}{c}{\textcolor{cadmiumgreen}{\cmark}} & \textcolor{cadmiumgreen}{\cmark} & \multicolumn{1}{c}{\textcolor{cadmiumgreen}{\cmark}} & \multicolumn{1}{c}{\textcolor{cadmiumgreen}{\cmark}} & \multicolumn{1}{c}{\textcolor{cadmiumgreen}{\cmark}} & \textcolor{cadmiumgreen}{\cmark}\\ \hline
\end{tabular}
}
\\
${}^\dagger$These designs are from ISCAS-89 benchmark suite, widely-used by the hardware security community~\cite{alkabani2007active,el2015integrated,yasin2017provably}.
\label{tab:overview_table}
\end{table*}

\subsection{Hardware Trojan Detection Techniques}
\label{sec:HT_detection_techniques}

Researchers attempt to detect HTs inserted during fabrication using (i)~logic testing~\cite{chakraborty2009mero,GA_SAT,TARMAC_TCAD,pan2021automated} and (ii)~side channel 
analysis~\cite{huang2016mers,HT_detection_cluster,HT_detection_fingerprinting,salmani2011layout,HT_detection_selfreferencing,jin2008hardware,HT_detection_chaos,HT_detection_eletromagnetic,lyu2021maxsense}. 
Both approaches apply input patterns to the underlying chip, and the only difference is the modality they monitor. 
In a logic testing-based approach, the defender monitors the output responses from the chip to detect deviations from the expected outputs 
using an HT-free (golden) chip~\cite{chakraborty2009mero,GA_SAT,TARMAC_TCAD,pan2021automated}. 
In a side channel-based approach, the defender monitors the physical characteristics (power consumed, path delay, electromagnetic emissions) of the chip to detect deviations from expected behavior~\cite{huang2016mers,HT_detection_cluster,HT_detection_fingerprinting,salmani2011layout,HT_detection_selfreferencing,jin2008hardware,HT_detection_chaos,HT_detection_eletromagnetic,lyu2021maxsense}.

\noindent\textbf{Limitations of Existing Techniques.} The aforementioned HT detection techniques suffer from key limitations, including: (i)~low success rate for detecting HTs~\cite{chakraborty2009mero,GA_SAT,huang2016mers}, (ii)~high algorithmic complexity~\cite{GA_SAT,huang2016mers,TARMAC_TCAD,lyu2021maxsense}, and (iii)~a large number of input patterns~\cite{chakraborty2009mero,GA_SAT,huang2016mers,TARMAC_TCAD,pan2021automated}, leading to increased test time and delayed deployment. 
However, the most pertinent drawback arises from a lack of proper security evaluation of these detection techniques, resulting from inappropriate/incorrect adversarial assumptions (\S\ref{sec:evaluation_methodology}).
The ramifications of an inappropriate security evaluation lead to a ``\textit{false sense of security}.''
To the best of our knowledge, despite more than a decade of research on detecting HTs (implanted during fabrication), there have been no concerted efforts to perform a
systematic and automated evaluation of HT detection techniques that can scale well to industrial-scale designs.

\subsection{Our Goals and Contributions}
\label{sec:goals_and_contributions}

In this work, we perform a systematic evaluation of HT detection techniques that aim to detect HTs
during the fabrication of ICs. 
We assume the role of a motivated and realistic adversary that inserts HTs to evade state-of-the-art HT detection techniques. 
However, detection techniques involve various algorithms ranging from (i)~random bit-flipping~\cite{chakraborty2009mero,huang2016mers}, (ii)~genetic algorithms~\cite{GA_SAT,lyu2021maxsense},
(iii)~graph-theoretic algorithms~\cite{TARMAC_TCAD},
to (iv)~reinforcement learning~\cite{pan2021automated}.
\textit{Therefore, we need a radical approach to systematically evaluate HT detection techniques by inserting stealthy HTs using an automated attack framework.}

However, several hurdles exist in designing an automated attack framework that inserts stealthy HTs during fabrication. 
First, the underlying design is a sea of gates and nets; it is computationally challenging to examine each net individually to create an HT. 
Second, the input patterns generated by the HT detection techniques are not deterministic. 
Third, an adversary does not have apriori information regarding the locations tested by the defender.
This moving target (\textit{i.e.,} the stochasticity of locations checked by the defender, non-deterministic input patterns, large design space exploration) makes it challenging for an adversary to evade detection techniques.

We address the aforementioned hurdles and develop a scalable attack framework,~\mytool,~using reinforcement learning (RL). 
RL has shown great promise in navigating unknown and uncertain problem spaces and finding optimal or near-optimal solutions, as in the case of fuzzing~\cite{RL_Fuzzing,RL_Fuzzing_USENIX}, Internet of Things security~\cite{RL_IoT,RL_iot_xiao}, and cyber security~\cite{RL_cyber,RL_cyber_adversarial}. 
Hence, we formulate the task of inserting HTs, which evade the detection techniques under the uncertainty of the input patterns, as an RL problem (\S\ref{sec:methodology}).
However, we must overcome several challenges to realize an automated, practical, and scalable RL agent: \cc{1}~dependence on detection techniques, \textit{i.e.,} reliance on input patterns from detection techniques, \cc{2}~expensive reward computations, \textit{i.e.,} evaluation of HTs generated during training, \cc{3}~inefficiencies of the agent's choices, \textit{i.e.,} inhibiting the agent from generating HTs that are unsuitable, \cc{4}~lack of scalability, \textit{i.e.,} insert HTs in large designs like \texttt{AES}, \texttt{GPS}, and \texttt{mor1kx} processor), and \cc{5}~lack of variety of HTs, \textit{i.e.,} generate a large corpus of HTs for an adversary to choose from.

We overcome the challenges of \cc{1}~reliance on input patterns and \cc{2}~expensive reward computations by characterizing the design before training. 
Characterization helps us compute the rewards quickly (up to $16\times$ faster than the na\"ive approach) (\S\ref{sec:offline_characterization}). 
To reduce the \cc{3}~inefficient choices made by the agent, we trim the actions available to the agent at different time steps depending on the present state of the agent (\S\ref{sec:trimming}). 
Finally, we overcome challenges about \cc{4}~scalability and \cc{5}~limited variety of HTs by carefully pruning the search space for the agent (\S\ref{sec:pruning}).

By solving these challenges, we develop an automated and scalable RL-based framework,~\mytool,~that performs a systematic evaluation of HT detection techniques by inserting stealthy HTs.
~\mytool~is agnostic to the choice of detection techniques considered in this work and scalable to practical designs, such as \texttt{AES}, \texttt{GPS}, and \texttt{mor1kx} processor. 
~\mytool~generated HTs evade eight HT detection techniques from logic testing (\S\ref{sec:evaluation_logic_testing}) and side channel-based detection approaches (\S\ref{sec:eval_side_channel_HTs}), and we showcase two case studies to demonstrate cross-layer, system-level attacks on \texttt{mor1kx} processor (\S\ref{sec:case_studies}).
The \textbf{contributions} of our work are as follows.

\begin{itemize}[leftmargin=*]

\item We develop an automated, scalable, and practical RL-enabled HT insertion framework,~\mytool~, that successfully evades several HT detection techniques, including the state-of-the-art.
To the best of our knowledge, our work is the first to use RL for developing a successful attack in supply chain security (\S\ref{sec:methodology}).

\item We demonstrate the generalization power of~\mytool~by evading eight HT detection techniques from logic testing and side channel categories (Table~\ref{tab:overview_table}). 
These techniques have been published in premier security venues, have been widely regarded in the industry, and cited in academia, and/or are state-of-the-art.
\mytool~achieves average attack success rates of $47\times$ and $211\times$ compared to randomly inserted HTs against state-of-the-art logic testing and side channel techniques (\S\ref{sec:evaluation_logic_testing} and \S\ref{sec:eval_side_channel_HTs}).

\item We showcase the efficacy of~\mytool~generated HTs on designs ranging from the widely used ISCAS-85 and ISCAS-89 benchmark suites to open-source \texttt{MIPS} and \texttt{mor1kx} processors, \texttt{AES}, and \texttt{GPS} (upto $\approx200,000$ gates; \S\ref{sec:experiments}).

\item We demonstrate how an adversary can repurpose~\mytool~to design HTs that not only evade detection, but also cause practical, cross-layer, real-world attacks through two case studies (privilege escalation and kill switch) on \texttt{mor1kx} processor (\S\ref{sec:case_studies}).

\item To foster further research in the area of HT detection and HT insertion techniques, we will open-source our codes and HT-infested designs.
Our developed HTs can be used by researchers to evaluate the efficacy of their detection techniques.

\end{itemize}
\section{Background and Preliminaries}
\label{sec:background}

We first provide an overview of hardware Trojans (HTs) and explain different HT detection techniques and their evaluation methodologies, followed by an introduction to reinforcement learning (RL).

\subsection{Hardware Trojans (HTs)}
\label{sec:hardware_trojans}

HTs are malicious logic inserted by adversaries to achieve a disruptive impact on ICs~\cite{agrawal2007trojan,hicks2010overcoming,sturton2011defeating,ICAS}.
HTs can (i)~operate as a hardware backdoor that leaks sensitive information such as cryptographic keys~\cite{lin2009trojan}, (ii)~cause denial-of-service during regular operation~\cite{baumgarten2011case}, and/or (iii)~effect a deviation in the functionality of the design~\cite{yang2016a2}.
Typically HTs can be inserted anywhere in the IC supply chain ranging from the register-transfer level (adversary is the third-party intellectual property provider) to the chip layouts (adversary in the foundry).
A taxonomy of HTs can be found in the Appendix~\ref{app:appendix_HT_taxonomy}.
An HT consists of two components: \textbf{trigger} and \textbf{payload}.
The trigger is the activation mechanism of an HT, and the trigger is activated by \textbf{rare nets} (\textit{i.e.,} nets whose logic values are strongly biased).
In particular, the trigger is activated when the rare nets assume their rare values.
These rare nets have an \textbf{activity probability} (\textit{i.e.,} the probability of a net being a $1$ or $0$) below a certain \textbf{rareness threshold}.
Once the trigger is excited, the payload gets activated, causing a malicious effect.
A \textbf{stealthy HT} must be (i)~malicious (jeopardize the chip's functionality or leak sensitive information or degrade the performance of the device) and (ii)~undetectable (would not be detected by any prior and state-of-the-art HT detection techniques).

\subsection{Prior Work on Hardware Trojan Detection} 
\label{sec:prior_work}

When the foundry is untrusted, HT detection techniques are classified into two broad categories: logic testing and side channel analysis.
Logic testing-based techniques detect HTs by applying test patterns to the HT-infested design to activate the trigger~\cite{chakraborty2009mero,TARMAC_TCAD,GA_SAT,pan2021automated}.
On the other hand, side channel-based detection techniques detect HTs by monitoring the deviations of the side-channel measurements (power consumption, path delay, electromagnetic emissions) of an HT-infested design from the expected measurements of a golden, \textit{i.e.,} HT-free, design~\cite{huang2016mers,lyu2021maxsense,HT_detection_cluster,HT_detection_fingerprinting,salmani2011layout,HT_detection_selfreferencing,jin2008hardware,HT_detection_chaos,HT_detection_eletromagnetic}.

Although there are many noteworthy HT detection techniques for both categories, we choose eight techniques for our evaluation, as explained next. 
Our selection spans from the earliest technique with industrial adoption, namely MERO~\cite{chakraborty2009mero}, to the latest one that uses an RL algorithm as a detection tool, namely TGRL~\cite{pan2021automated}. 
We also select techniques that have a high impact (measured through citations) from a diverse set of conferences/journals: (i)~MERS~\cite{huang2016mers} from ACM CCS, (ii)~TARMAC~\cite{TARMAC_TCAD}, an industry-adopted technique~\cite{mishra2021trigger_patent}, from IEEE TCAD, a top computer-aided system design transactions, (iii)~GA+SAT~\cite{GA_SAT} from CHES, a top hardware security conference, and (iv)~MaxSense~\cite{lyu2021maxsense}, the state-of-the-art technique for side channel-based HT detection (62$\times$ better than MERS~\cite{huang2016mers}).
These techniques claim the following properties: (i)~scalability, (ii)~efficiency, (iii)~accuracy in detecting stealthy HTs, and (iv)~ease of integration with IC design tools and flow. 
In addition to MERS and MaxSense, there are a few other high-impact works on detecting HTs using side-channel analysis, such as ~\cite{jin2008hardware,salmani2011layout,wei2011scalable,banga2008region}.
However, since MERS and MaxSense magnify the impact of all of these techniques, we chose MERS~\cite{huang2016mers} and MaxSense~\cite{lyu2021maxsense}.

\noindent\textbf{Logic testing-based Techniques.}
\textbf{MERO} generates test patterns that activate each rare net $K$ times~\cite{chakraborty2009mero}.
The hypothesis is that if all the rare nets are activated $K$ times, the resultant test patterns are likely to activate unknown triggers. 
\textbf{GA+SAT} generates test patterns by utilizing a genetic algorithm and Boolean satisfiability (SAT)~\cite{GA_SAT}.
\textbf{TARMAC} generates test patterns using clique enumeration~\cite{TARMAC_TCAD}.
\textbf{TGRL} uses RL to generate a set of test patterns to maximize the likelihood of activating HTs~\cite{pan2021automated}.

Researchers evaluate the efficacy of logic testing-based detection techniques using \textbf{HT activation rate}~\cite{chakraborty2009mero,GA_SAT,TARMAC_TCAD,pan2021automated}, defined as the percentage of HTs activated by the test patterns. 
Mathematically, the HT activation rate is $\left(\frac{\text{Number of HTs activated}}{\text{Total number of HTs inserted}}\right) \times 100 \%$.

\noindent\textbf{Side channel-based Techniques.} \textbf{MERS} extends the idea of MERO to side-channel metrics by generating test patterns that cause rare nets to switch from their non-rare to rare values $K$ times~\cite{huang2016mers}.
Doing so increases the likelihood of activating an HT, which in turn increases the deviation of the measured side-channel from expected (\textit{i.e.,} golden) values.
\textbf{MERS-h} reorders test patterns generated by MERS to simultaneously maximize the activity in rare nets and minimize the activity in non-rare nets.
Unlike MERS-h, \textbf{MERS-s} relies on the actual behavior of the golden circuit (obtained through functional simulations) to measure the activities in the rare and non-rare nets.
\textbf{MaxSense} exploits input affinity to generate test patterns that maximize switching in the malicious logic while minimizing switching in the rest of the circuit~\cite{lyu2021maxsense}.

Researchers evaluate the efficacy of side channel-based detection techniques using \textbf{side-channel sensitivity}~\cite{huang2016mers,lyu2021maxsense}, which measures the amount of switching caused in an HT-infested design relative to the switching caused in an HT-free design.
Let $G$ be the golden design (\textit{i.e.,} HT-free), $HT$ be the HT-infested design, and $(u_i,v_i)$ denote the $i^{\text{th}}$ pair of consecutive test patterns, then, the side-channel sensitivity of the HT-infested design $HT$ is
\begin{equation*}
    sensitivity_{HT} = \max_{(u_i,v_i)} \left( \frac{|switching_{(u_i,v_i)}^{HT} - switching_{(u_i,v_i)}^{G}|}{switching_{(u_i,v_i)}^{G}} \right)
\end{equation*}
We use these metrics (\textbf{HT activation rate} and \textbf{side-channel sensitivity}) to showcase the efficacy of~\mytool.

\subsection{Evaluation Methodology of Detection Techniques}
\label{sec:evaluation_methodology}

The aforementioned  detection techniques span over a decade---from $2009$ to $2021$.
Collectively, they cover several algorithms, ranging from random bit-flipping~\cite{chakraborty2009mero}, genetic algorithm~\cite{GA_SAT,lyu2021maxsense}, graph-theoretic algorithm~\cite{TARMAC_TCAD},  reinforcement learning-based algorithm~\cite{pan2021automated}, to side channel-based detection~\cite{huang2016mers,lyu2021maxsense}. 
However, all these techniques lack proper evaluation in two aspects.

Firstly, all aforementioned detection techniques assume that an adversary randomly samples HTs from the rare nets.
As a result, these detection techniques claim very high efficacy ($>90\%$ HT activation rate).
Similarly, benchmark suites of HT-infested designs (e.g., TrustHub~\cite{TrustHub,salmani2013design} and~\cite{yu2019improved}) consist of randomly inserted HTs.
However, this assumption does not reflect the real-world scenario since an adversary in a foundry is not constrained to insert HTs randomly, as evidenced in~\cite{sturton2011defeating,skorobogatov2012breakthrough,becker2013stealthy,yang2016a2}.
On the contrary, an adversary inserts HTs such that the likelihood of detection is minimal. 
Our results demonstrate that the HTs generated by~\mytool~cause a drastic reduction in the efficacy of logic testing-based techniques (Table~\ref{tab:results_logic_testing}) and side channel-based techniques (Table~\ref{tab:results_side_channel}).

Secondly, all aforementioned detection techniques consider designs that contain just a few thousand gates.
Such designs are not practical since modern design intellectual property cores contain at least a few hundred thousand gates (e.g., \texttt{AES}, \texttt{GPS}, \texttt{mor1kx} processor). 
Furthermore, detecting HTs in designs containing a few thousand gates is relatively easier since there are limited places where an HT can be inserted.
Thus, the evaluation methodology used by the state-of-the-art HT detection techniques is inappropriate, and it gives a \textit{``false sense of security.''}
We use RL as a litmus test to demonstrate these issues and obtain the actual efficacy of HT detection techniques.

\subsection{Reinforcement Learning (RL)}
\label{sec:RL}
 
RL is a machine learning technique where an agent learns how to act in an (unknown) environment through actions and receives feedback through rewards.
Unlike supervised/unsupervised learning, which requires labeled/unlabeled data; RL does not require pre-defined data. 
RL is used to solve problems involving an optimal sequence of decisions by modeling the underlying problem as a 
Markov decision process~\cite{sutton_barto_reinforcement,wiering2012reinforcement}.
While many fields such as software fuzzing~\cite{RL_Fuzzing,RL_Fuzzing_USENIX}, Internet-of-Things security~\cite{RL_IoT,RL_iot_xiao}, and cyber security\cite{RL_cyber,RL_cyber_adversarial} have reaped the benefits of using RL, hardware security is still in its infancy to reap the powers of RL.
\section{Threat Model}
\label{sec:threat_model}

Before diving into the specifics regarding~\mytool, we outline the location, capabilities, and goal of an adversary.

\noindent\textbf{Adversary Location.} We assume the adversary is present in the untrustworthy foundry.
Our threat model selection is motivated by the fact that most IC design companies outsource fabrication to overseas foundries~\cite{fabless} (\S\ref{sec:introduction}).
In fact, the U.S. \textit{Department of Defense} 2022 action plan for securing defense-critical supply chains points out that 88\% of microelectronic fabrication is performed overseas, representing a substantive security threat~\cite{DoD_supply_chain}.

\noindent\textbf{Adversarial Capabilities.} We outline the capabilities of an adversary 
consistent with state-of-the-art research in hardware Trojans (HTs)~\cite{chakraborty2009mero,GA_SAT,TARMAC_TCAD,pan2021automated,huang2016mers,lyu2021maxsense}.

\begin{enumerate}[wide, labelwidth=!, labelindent=0pt]

\item [1.] An adversary obtains the gate-level design by reverse-engineering the Graphics Database System II (GDSII).\footnote{GDSII is a database file used for exchanging IC layout information.}
The adversary has access to and know-how regarding state-of-the-art reverse engineering equipment~\cite{reverse_tag,chipworks,Degate}.
\item [2.] An adversary can construct the trigger using only rare nets. 
To that end, they can compute the list of rare nets by performing functional simulations using any tool (academic or commercial).
\item [3.] An adversary has resources (placement sites for trigger and payload and routing resources for connecting wires) available in the GDSII to insert the trigger and payload. 
\item [4.] An adversary does not know the input patterns used by the defender for post-fabrication testing. 
We assume that an adversary knows the type of HT detection technique(s) used by the defender.

\end{enumerate}

\noindent\textbf{Adversarial Goal.} The goal of the adversary is to cause a disruptive impact (e.g., leak secret information like cryptographic keys~\cite{lin2009trojan}, perform privilege escalation~\cite{sturton2011defeating,yang2016a2}) on the ICs by inserting HTs while evading all detection techniques employed by the defender.
Our work focuses on additive HTs since all considered detection techniques assume additive HTs~\cite{chakraborty2009mero,GA_SAT,TARMAC_TCAD,pan2021automated,huang2016mers,lyu2021maxsense,ICAS}.
\section{\mytool: Attacking Hardware Trojan Detection Techniques}
\label{sec:methodology}

We initially explain why reinforcement learning (RL) is suited for our problem definition and our preliminary formulation; however, our preliminary formulation has several limitations.
Subsequently, we explain the challenges and the steps we took to overcome them, and finally we outline the architecture for~\mytool.

\begin{figure}[tb]
\centering
\includegraphics[trim=1cm 0.9cm 0.5cm 0.6cm, clip, width=0.4\textwidth]{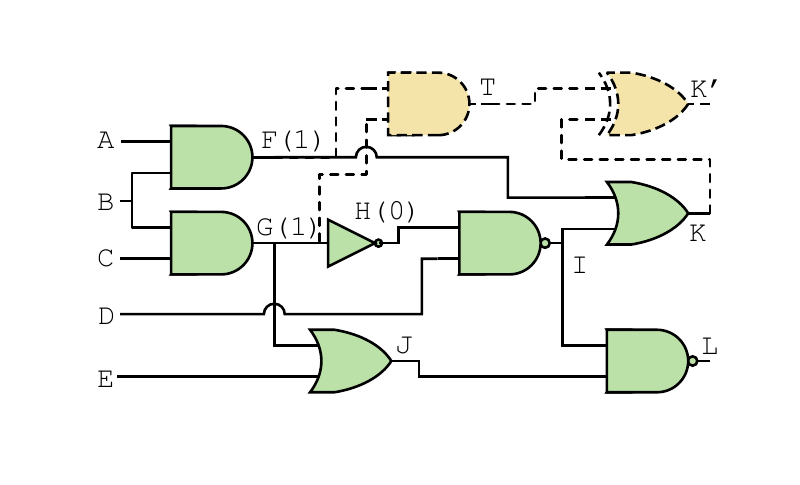}
\smallerspacecaption
\caption{The trigger \texttt{T} (constructed from \texttt{F} and \texttt{G}), when activated, triggers the payload to flip the value of \texttt{K}.}
\label{fig:Trojan}
\end{figure}

\subsection{Why Reinforcement Learning?}
\label{sec:why_RL}

Inserting stealthy hardware Trojans (HTs) (\textit{i.e.,} evading HT detection techniques) necessitates the judicious selection of trigger nets (rare nets that constitute the trigger).
A concerted selection strategy of trigger nets has two features, as explained next.

\noindent\textbf{Sequential Decision Making.} If an adversary wants to construct the trigger using $\alpha$ rare nets, a na\"ive approach would be to select all $\alpha$ nets simultaneously.
However, since nets are part of the design, their logic values are dependent on each other.
Thus, na\"ively selecting $\alpha$ rare nets would be sub-optimal.
For instance, consider the design in Figure~\ref{fig:Trojan}, where there are three rare nets (\texttt{F}, \texttt{G}, \texttt{H} with rare values \texttt{1}, \texttt{1}, and \texttt{0}, respectively).
However, the value of \texttt{H} depends on the value of \texttt{G}.
In fact, whenever \texttt{G} takes its rare value, \texttt{H} also takes its rare value.
Thus, selecting both these nets would be sub-optimal because any test pattern that activates \texttt{G} will also activate \texttt{H}, leading to easy detection.
Alternatively, depending on the design's structure (\textit{i.e.,} connectivity among logic gates), selecting one particular rare net can lead to a non-selection of other rare nets to form a valid HT (an HT that can be activated).
Hence, a better approach would be to 
choose the rare nets sequentially, \textit{i.e.,} select one rare net, understand its impact on the other rare nets in the design, and then select the next rare net.
In other words, a sequence of optimal decisions is required to construct an HT that evades detection techniques.

Secondly, there is an \textbf{uncertainty in the test patterns} generated by the detection techniques, \textit{i.e.,} the potential triggers they check for are \textit{unknown to an adversary.}
For instance, in a design with $1000$ rare nets, constructing an HT with four rare nets can be done in ${}^{1000}C_4 \approx 4.1\times10^{10}$ ways. 
However, an adversary does not know the combinations checked by the defender.
Hence, a concerted strategy should construct HTs (\textit{i.e.,} select trigger nets) under the uncertainty of the locations checked by the detection techniques.

Problems with these two features (\textbf{sequential decision making} and \textbf{uncertainty in the environment}) are the kind of tasks that RL algorithms solve well---making an optimal sequence of decisions under uncertainty~\cite{sutton_barto_reinforcement,wiering2012reinforcement}. 
Hence, we formulate the HT generation problem as an RL problem.

\subsection{Preliminary Formulation}
\label{sec:prelim_formulation}

We now explain our preliminary formulation. 
We assume that an adversary constructs a trigger using $T_{wid}$ rare nets,\footnote{The number of rare nets used to construct the trigger is defined as \textbf{trigger width}.} throughout this section.
Recall that, to evade existing detection techniques; we need to construct the trigger out of those combinations of rare nets that are not simultaneously activated by test patterns from existing detection techniques (\S\ref{sec:why_RL}). 
Note that if an adversary knew the \textit{exact test patterns that a defender used to detect HTs}, the solution is straightforward---generate a trigger from rare nets not activated by the test patterns.

Since an adversary knows the HT detection technique(s) used by the defender (\S\ref{sec:threat_model}), 
they generate a set of test patterns, which will be different from the patterns used by the defender.
Given these test patterns, an adversary aims to generate triggers (and hence HTs) that are unlikely to be detected by the defender. 
We construct the HT triggers by mapping the HT generation problem as an RL problem, as described below.

\begin{itemize}[leftmargin=*]

\item \textbf{States $\mathcal{S}$} is the set of all subsets of rare nets. 
An individual state $s_t$ represents the set of rare nets chosen by the agent at time $t$.

\item \textbf{Actions $\mathcal{A}$} is the set of all rare nets. An individual action $a_t$ is the rare net chosen by the agent at time $t$.

\item \textbf{State transition $\mathcal{P}(s_{t+1}|a_t,s_t)$} is the probability that action $a_t$ in state $s_t$ leads to the state $s_{t+1}$.
If the chosen rare net (\textit{i.e.,} the action) is \textbf{compatible}\footnote{A set of rare nets is compatible if an input pattern exists that can activate all the rare nets in the set simultaneously. 
Alternatively, we say that a rare net, $r_i$, is compatible with a given set of rare nets, $R$, if an input pattern exists that can activate $r_i$ and all nets in $R$ simultaneously.} with the current set of rare nets (\textit{i.e.,} the current state), we add the chosen rare net to the set of compatible rare nets (\textit{i.e.,} the next state). 
Otherwise, the next state remains the same as the current state. 
Thus,
    \[
    s_{t+1}=
    \begin{dcases}
        \{a_t\} \cup s_t, & \text{if } a_t\text{ is compatible with }s_t\\
        s_t,              & \text{otherwise}
    \end{dcases}
    \]

\item \textbf{Reward function $\mathcal{R}(s_t,a_t)$.} Denote $\mathcal{I}_\mathsf{C}: \mathcal{S}\times\mathcal{A} \rightarrow \{0,1\}$. Given any state action pair $(s,a)$, $\mathcal{I}_\mathsf{C}(s,a)$ indicates whether $a$ is compatible with $s$ or not. Denote $\mathcal{I}_{A}^{TP}: \mathcal{S} \rightarrow \{0,1\}$. Given a state $s$, $\mathcal{I}_{A}^{TP}(s)$ indicates if $s$ is activated by some test pattern in $TP$ or not. Then, the reward function is defined as
    \[
    \mathcal{R}(s_{t},a_t)=
    \begin{dcases}
        0,&\text{if } (\left|s_{t+1}\right| \neq T_{wid}) \ \land\ (\mathcal{I}_\mathsf{C}(s_t,a_t)=1)\\
        \rho_1,&\text{if } (\left|s_{t+1}\right| \neq T_{wid})\ \land\ (\mathcal{I}_\mathsf{C}(s_t,a_t)=0)\\
        0,&\text{if } (\left|s_{t+1}\right| = T_{wid})\ \land\ (\mathcal{I}_{A}^{TP}(s_{t+1})=1)\\
        \rho_2,&\text{if } (\left|s_{t+1}\right| = T_{wid})\ \land\ (\mathcal{I}_{A}^{TP}(s_{t+1})=0)
    \end{dcases}
    \]
    Note that $\rho_1<0$, \textit{i.e.,} a penalty, and $\rho_2>0$. 
    Also, $TP$ denotes a set of test patterns from a detection technique (e.g., TGRL~\cite{pan2021automated}).
    We design the reward function so that the agent tries to construct triggers that are not activated by the generated test patterns.
    \item \textbf{Discount factor $\gamma$} $(0 \leq \gamma \leq 1)$ indicates the importance of future rewards relative to the current reward.
\end{itemize}

\noindent In the training process, we initialize the agent with a random rare net, and it chooses other rare nets at each step.
At the end of the episode (\textit{i.e.,} after $T_{wid}-1$ steps), the rare nets chosen by the agent (so far) reflect the generated HT.

\noindent\textbf{Challenges.} Although this preliminary formulation achieves a high success rate in evading HT detection techniques (e.g., on average, our RL-generated HTs are $49.87\times$ more stealthy against TARMAC than random HTs for \texttt{c6288}, \texttt{c7552}, \texttt{s13207}, and \texttt{s15850},\footnote{Widely used designs by hardware security community~\cite{alkabani2007active,el2015integrated,yasin2017provably}.} it faces the following challenges: \cc{1}~it relies on test patterns from detection techniques, \cc{2}~simulations for reward computation are expensive, leading to large training time for medium-sized designs (e.g., agent does not generate stealthy HTs for \texttt{MIPS} with $\approx~25,000$ gates even after $12$ hours of training),
\cc{3}~it is inefficient in choosing rare nets because the agent chooses some rare nets that are incompatible,
\cc{4}~it is not scalable to large designs like processor cores or crypto-engines (e.g., \texttt{mor1kx} and \texttt{AES}), and \cc{5}~it generates only $67$ stealthy HTs for \texttt{mor1kx},\footnote{An adversary would prefer to have a corpus of stealthy HTs since some HTs might be infeasible for insertion due to resource constraints (lack of placement sites for trigger and/or lack of routing resources) or cause noticeable changes in side channels, thereby facilitating detection.}
thus limiting the options for an adversary.
Next, we explain these challenges and how we overcome them.

\subsection{Offline Characterization}
\label{sec:offline_characterization}

\textbf{Challenge \cc{1}.} A fundamental limitation of our preliminary formulation is that it relies on test patterns from a detection technique like TGRL~\cite{pan2021automated} or TARMAC~\cite{TARMAC_TCAD} for training. 
Ideally, the RL agent should be agnostic to the test patterns, thereby agnostic to the HT detection technique.
If the agent relies on test patterns and researchers develop a better detection technique, the agent would need to be retrained with test patterns from this new technique. 
However, retraining is time-consuming since one needs to do it for every new technique. 
Hence, we aim to develop an agent that is agnostic to test patterns from detection techniques so that the agent can be used against other techniques in the future.

\noindent\textbf{Challenge \cc{2}.} Another reason for devising the agent to be agnostic to test patterns is time complexity for reward computation.
In our preliminary formulation, computing the reward requires us to check if the final state of the agent, \textit{i.e.,} the trigger consisting of the set of $T_{wid}$ rare nets, is activated by any test patterns in $TP$.
To that end, we need to (i)~construct the trigger, (ii)~append it to the original design, and (iii)~simulate the design using all test patterns in $TP$.
Performing these steps is computationally expensive. 
Asymptotically, the time complexity of simulating a design having $g$ gates with $p$ test patterns is $\mathcal{O}(gp)$.
Although theoretically, the complexity is linear, practically, each reward computation takes several seconds even for medium-sized designs (e.g., for \texttt{MIPS} $\approx~25,000$ gates, the rate is $\approx13$ seconds/reward computation), and RL agents require several thousand episodes for convergence~\cite{sutton_barto_reinforcement}. 
Hence, the training time required for such an approach is impractical for designs with tight time-to-market constraints in the semiconductor industry.

\noindent\textbf{Solution 1.} To \cc{1}~avoid reliance on test patterns and \cc{2}~reduce the time required to compute the reward, we modify the training process by characterizing the target design (before training) and saving the intermediate results.
The intermediate results are used during training for computing the reward quickly.
Next, we explain how we refine the preliminary formulation by modeling the goal of an adversary using \textit{offline characterization}.

To increase the likelihood of evading unknown test patterns, an adversary needs to \textit{identify combinations of rare nets that are compatible but are least likely to be activated simultaneously.} 
However, identifying combinations of rare nets is non-trivial as the search space is huge:
assuming $N$ rare nets in a design, and $T_{wid}$ being four, there are ${}^NC_4$ combinations, which increases combinatorially with the number of rare nets.
For example, \texttt{MIPS} has $1,005$ rare nets leading to $4.2\times 10^{10}$ different triggers.
The problem is further exacerbated because picking a set of rarest $T_{wid}$ nets might not work since (i)~they are not compatible (\textit{i.e.,} the constructed HT trigger is invalid as it can never be activated), or (ii)~they are very likely to be activated by detection techniques. 
Thus, to estimate the likelihood of compatible rare nets being activated simultaneously, we leverage a randomized algorithm, which we describe next.

\begin{table}[tb]
\caption{Comparison of training rates for online and offline reward calculation methods for \texttt{MIPS}.}
\label{tab:online_vs_offline_comp}
\begin{tabular}{ccc} \toprule
\multirow{2}{*}{\begin{tabular}[c]{@{}c@{}}Reward calculation\\ method\end{tabular}} & 
\multicolumn{2}{c}{Training rate}  \\ \cmidrule(lr){2-3}
& steps/min & episodes/min \\ \midrule
Online reward & $\approx13.88$ & $\approx4.63$ \\ 
Offline reward & $\approx234.82$ & $\approx78.23$ \\ 
\textbf{Speedup} & $\mathbf{16.91\times}$ & $\mathbf{16.89\times}$ \\ \bottomrule
\end{tabular}
\end{table}

The algorithm runs for $T$ iterations and maintains $T+1$ sets, $\mathcal{C}_1,\mathcal{C}_2,\ldots,\mathcal{C}_T$ ($\mathcal{C}_i$ denotes the compatible rare nets in the $i^\text{th}$ iteration), and a set of unexplored rare nets $\mathcal{U}$.
At the start of $i^{\text{th}}$ iteration, $\mathcal{C}_i$ and $\mathcal{U}$ are initialized as an empty set and as a set of all rare nets in the design, respectively. 
Next, a random rare net, $r_j$, is selected from $\mathcal{U}$. 
If $r_j$ is compatible with $\mathcal{C}_i$,
it is removed from $\mathcal{U}$ and added to $\mathcal{C}_i$; otherwise, $r_j$ is just removed from $\mathcal{U}$.
This process is repeated until $\mathcal{U}$ is empty, \textit{i.e.,} $N$ times.
Then, the algorithm starts the next iteration, $i+1$.\footnote{Note that each iteration of this algorithm is independent. 
Hence, in our implementation, we 
parallelize the algorithm.}
At the end of $i^\text{th}$ iteration, we get $\mathcal{C}_i$, a set of compatible rare nets. 
These $\mathcal{C}_i$'s, $\forall i \in \{1,2,\ldots,T\}$, are used to compute the likelihood of a given set of compatible rare nets being activated simultaneously, which is translated into a reward for the agent.
More specifically, the reward function is updated as
    \[
    \mathcal{R}(s_{t},a_t)=
    \begin{dcases}
        0,&\text{if } (\left|s_{t+1}\right| \neq T_{wid}) \ \land\ (\mathcal{I}_\mathsf{C}(s_t,a_t)=1)\\
        \rho_1,&\text{if } (\left|s_{t+1}\right| \neq T_{wid})\ \land\ (\mathcal{I}_\mathsf{C}(s_t,a_t)=0)\\
        0,&\text{if } (\left|s_{t+1}\right| = T_{wid})\ \land\ (\exists~i \mid s_{t+1} \subseteq \mathcal{C}_i)\\
        \rho_2,&\text{if } (\left|s_{t+1}\right| = T_{wid})\ \land\ (\nexists~i \mid s_{t+1} \subseteq \mathcal{C}_i)
    \end{dcases}
    \]
Following this reward, the agent tries to select rare nets that are compatible but least likely to be activated simultaneously and, thereby, likely to evade unknown test patterns from detection techniques.
Empirical results on designs of different types and sizes validate this hypothesis (\S\ref{sec:experiments}).

Since the characterization is performed before training, the time complexity of computing rewards using this approach is $\mathcal{O}(T)$, where $T$ is the number of sets ($\mathcal{C}_i$) calculated during characterization. 
The value of $T$ controls the trade-off between runtime and the efficacy of the generated HTs. 
Larger (smaller) values of $T$ require larger (less) time for characterization and reward computation and generate HTs that are more (less) likely to evade the detection techniques.
Note that the theoretical complexity of reward calculation during training is independent of the size (\textit{i.e.,} the number of gates) of the underlying design. 
In practice, each reward computation takes a few milliseconds, even for large designs such as \texttt{AES}, \texttt{GPS}, and \texttt{mor1kx} with more than $150,000$ gates.

Thus, the offline characterization approach helps alleviate the reliance on test patterns from HT detection techniques. It also helps reduce the reward computation time, leading to faster training. 
Table~\ref{tab:online_vs_offline_comp} depicts the training rates (in steps/minute and episodes/minute) for the two approaches for computing rewards: (i)~using test patterns during training and (ii)~using the information saved during offline characterization. 
Note that using offline characterization to pre-compute information and using them during training is more than $16\times$ faster than the online calculation approach.

\subsection{Trimming: Avoiding Redundant Actions}
\label{sec:trimming}

\textbf{Challenge \cc{3}.} Recall that, in the preliminary formulation, the actions available to the agent (at each time step) remain the same irrespective of the state of the agent.
This leads to situations where the agent chooses an action that (i)~has already been chosen in the past (\textit{i.e.,} a rare net that is already present in the current state $s_t$) or
(ii)~is known to not lead to a new state (\textit{i.e.,} a rare net known to be incompatible with at least one of the rare nets in the current state).
Choosing such actions makes the training process inefficient as these actions do not lead the agent to a new state. 
Consequently, the time spent by the agent on such steps is wasted.

\noindent\textbf{Solution 2.} To increase the efficiency of the agent in choosing actions while training (thereby reducing the training time), we trim the actions available to the agent based on the state at any given time step. 
Doing so increases the likelihood that at each time step, the agent chooses an action that leads it to a new state.

Ideally, to perform trimming during training, we would like to know which combinations of rare nets are compatible (or not).
Therefore, we compute the (in)compatibility of a set of rare nets using a SAT solver. 
However, if there are $N$ rare nets, we would need to invoke the SAT solver $2^N$ times ($2^N$ subsets of $N$ rare nets) to compute the compatibility/incompatibility of all possible subsets of rare nets.
Since such an approach is infeasible, we use an approximate approach, as explained next.

Instead of computing the compatibility of all subsets of rare nets, we only compute the compatibility of all subsets of size up to $S~(<<N)$.
This compatibility information is computed once and saved for later use.\footnote{Since the compatibility computation for each unique pair is independent, we parallelize the computation to reduce the runtime. 
Additionally, since we already generate this compatibility information for trimming actions, we also use it in the offline characterization algorithm (\S\ref{sec:offline_characterization}) when we check if a random rare net $r_j$ from $\mathcal{U}$ is compatible with $\mathcal{C}_i$. 
This helps reduce the runtime of the characterization algorithm.}
During training, when the episode begins in a random state (\textit{i.e.,} a singleton set with a random rare net), all actions (\textit{i.e.,} rare nets) that are not compatible with the initial state are trimmed off. 
Then, at each step, after choosing a valid action, the available action list is updated to trim off all rare nets that are not compatible with the latest chosen action.

\subsection{Scalability: Pruning Number of Branches}
\label{sec:pruning}

Incorporating solutions 1 and 2 enables the agent to train well and generate stealthy HTs for designs
such as \texttt{MIPS} ($\approx 25,000$ gates). 
However, the agent's performance on 
large designs (\texttt{AES}, \texttt{GPS}, and \texttt{mor1kx} with $> 150,000$ gates) is sub-optimal, as explained next.

\begin{figure}[tb]
\includegraphics[trim=0.4cm 0.cm 0.cm 0.cm, clip,width=0.475\textwidth]{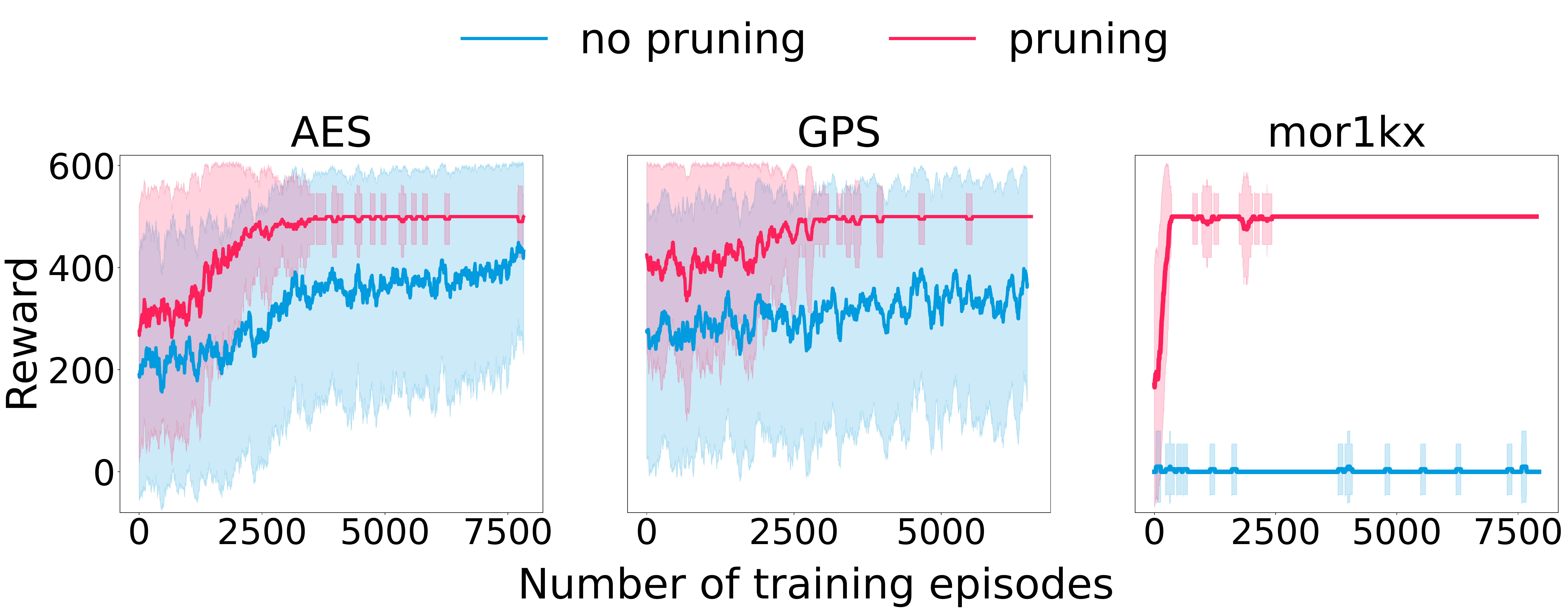}
\smallerspacecaption
\smallerspacecaption
\smallerspacecaption
\caption{Agent's inability/ability to learn without/with pruning. 
Total runtime for all plots is 12 hours.
The shaded region represents the standard deviation.}
\label{fig:demonstration_of_bad_and_good_reward_curves}
\end{figure}

\begin{figure*}[tb]
\centering
\includegraphics[trim=2.3cm 0.2cm 0.2cm 0.2cm, clip, width=\textwidth]{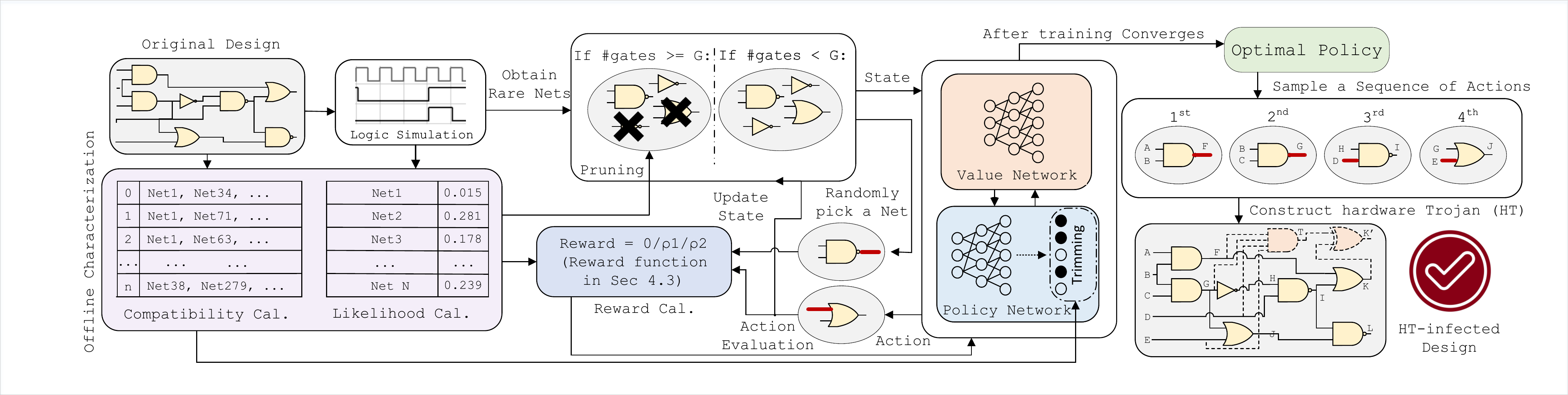}
\smallerspacecaption
\smallerspacecaption
\smallerspacecaption
\smallerspacecaption
\caption{Architecture of~\mytool.}
\label{fig:final_architecture}
\end{figure*}

\noindent\textbf{Challenge \cc{4}.} Our updated RL framework (including solutions 1 and 2) is not scalable to large designs such as \texttt{AES}, \texttt{GPS}, and \texttt{mor1kx}. 
For these designs, the agent is unable to learn an optimal/near-optimal policy even after $12$ hours of training (see the blue curve in Figure~\ref{fig:demonstration_of_bad_and_good_reward_curves}).
A possible workaround would be to train for longer time, however, such an approach is not scalable for designs with tight time-to-market constraints in the semiconductor industry.

\noindent\textbf{Challenge \cc{5}.} Another limitation with the training process is that for \texttt{mor1kx}, the agent generates only a small number of stealthy HTs.
This is due to the structure of the design, as described below.
Many rare nets are compatible with several other rare nets making it relatively easier for detection tools to detect HTs.
The agent generated only $67$ stealthy HTs across $12$ hours of training. 
Ideally, an HT generation framework should generate a large corpus of stealthy HTs, providing an adversary with several options (footnote 5).
Next, we outline how we overcome these challenges with a unified solution that aids the agent in 
choosing better actions.

\noindent\textbf{Solution 3.} Recall that we initialize the agent in a random state (\textit{i.e.,} with a random rare net).
The agent needs to learn an optimal sequence of decisions starting from each of the rare nets, \textit{i.e.,} the agent needs to learn $N$ branches, where $N$ is the number of rare nets in the design. 
Since \texttt{AES}, \texttt{GPS}, and \texttt{mor1kx} have a large number of gates, in $12$ hours, the agent performs only $7,923$, $6,562$, and $8,043$ episodes (\textit{i.e.,} trials), respectively. 
Thus, on average, for each branch, the agent only performs $9.27$, $7.69$, and $12.54$ episodes, respectively, which is insufficient for the agent to learn. 
A simple approach to increase the number of episodes per branch would be to increase the training time. 
However, this approach is neither feasible nor scalable. 
A better approach is to reduce, \textit{i.e.,} prune the number of branches for the agent to learn.
However, pruning needs to be done carefully so that \textit{the branches that contain stealthy HTs are not pruned off.}

We rely on the data collected during offline characterization to decide which branches (\textit{i.e.,} rare nets as initial states) to prune. 
We estimate the likelihood of each branch leading to easy-to-detect HTs and choose the branches that are least likely. 
The likelihood, $L(\cdot)$, of each branch (\textit{i.e.,} rare net, denoted as $r_j \text{ for } j = 1, 2, \ldots, N$) is estimated as the number of times that rare net is activated in compatible sets, $\mathcal{C}_i$'s, computed during characterization. 
Mathematically, $L(r_j) = \left|\{i \mid r_j \in \mathcal{C}_i\}\right|$.
$B$ branches with the lowest $L(\cdot)$ are picked, and the other branches are pruned off.

$B$ controls the trade-off between the (ease of) learning of the agent and the quality of generated HTs. 
Large values of $B$ provide the agent a large space of candidate HTs to choose from.
However, the agent would need to learn more branches, requiring more time. 
Smaller values of $B$ restrict the space from where the agent can choose HTs (\textit{i.e.,} it could potentially miss out on some stealthy HTs from the pruned-off branches), but assuming a fixed training time, the number of episodes that the agent spends on each of the $B$ branches would be larger, allowing better learning.
The red curves in Figure~\ref{fig:demonstration_of_bad_and_good_reward_curves} showcase the reward trends obtained after implementing pruning. 
After pruning, the reward curves ramp up quickly, \textit{i.e.,} the agent learns to generate stealthy HTs. 
Additionally, we observe that with pruning, the standard deviation in the reward values reduces to $0$ as the agent learns, whereas, without pruning, even after $12$ hours (\textit{i.e.,} $>6,000$ episodes), either the standard deviation is still very large ($\approx 170$ for \texttt{AES} and $\approx 214$ for \texttt{GPS}) or the reward is close to $0$ (for \texttt{mor1kx}).
Having overcome all these challenges, we discuss the final architecture of~\mytool~next.

\begin{table}[tb]
\caption{Comparison of HT attack success rate (in percentage; defined as $\mathbf{100}$ $-$ HT activation rate) against TARMAC for \texttt{AES}, \texttt{GPS}, and \texttt{mor1kx}: Random vs.\ RL without pruning vs.\ RL with pruning ($B=1$).}
\label{tab:init_pruning}
\resizebox{0.45\textwidth}{!}{
\begin{tabular}{ccccc} \toprule
\multirow{2}{*}{Design} & \multirow{2}{*}{\begin{tabular}[c]{@{}c@{}}Randomly-\\generated HTs\end{tabular}} & \multicolumn{3}{c}{RL-generated HTs} \\ \cmidrule(lr){3-5}
& & No pruning & With pruning & Improv. \\ \midrule
\texttt{AES} & $48$ & $81$ & $99$ & $\textbf{1.22}\times$\\ 
\texttt{GPS} & $62$ & $77$ & $95$ & $\textbf{1.23}\times$\\ 
\texttt{mor1kx} & $1$ & $9$ & $87$ & $\textbf{9.66}\times$\\ \bottomrule
\end{tabular}
}
\end{table}

\subsection{Final Architecture}
\label{sec:final_architecture}

Figure~\ref{fig:final_architecture} illustrates the final architecture of~\mytool.
Given a design, we first identify the rare nets (based on a rareness threshold) using functional simulations. 
Next, we perform two offline computations: (i)~compatibility information of each unique pair of rare nets (for trimming actions) and (ii)~likelihood of activation of different combinations of rare nets (\textit{i.e.,} offline characterization). 
Both computations are parallelized.
Then, we begin the training phase. 
If the design contains less than $\mathcal{G}$ gates, we initiate each training episode with a random rare net. 
If the design contains more than or equal to $\mathcal{G}$ gates, we use pruning (\S\ref{sec:pruning}) and initiate each training episode with a randomly picked rare net from the $B$ rare nets that are not pruned off.
In both cases, the agent takes action (\textit{i.e.,} chooses a rare net) according to the policy (parameterized by a neural network) and the trimmed action space (\textit{i.e.,} compatible rare nets). 
The action chosen by the agent is evaluated to compute the reward and to move the agent to the next state. 
This process repeats $T_{wid} -1$ times (because the first rare net is already provided to the agent as the initial state), constituting an episode. 
After a certain number of episodes, the underlying RL algorithm translates the rewards into losses, which are used to update the policy and value neural networks. 
Eventually, the parameters of the neural networks are tuned, losses become negligible, and reward saturates at the maximum value. 
Finally, we use the best episodes (\textit{i.e.,} episodes that have the largest rewards after the agent has converged) generated by the agent to create our HTs. 
\mytool~(i)~is independent of test patterns from HT detection techniques, (ii)~has inexpensive reward calculation while training, (iii)~selects rare nets effectively, and (iv)~is scalable to large designs.

The final architecture described above is independent of detection techniques and test patterns. 
The objective of the RL agent is not to evade a given set of test patterns; instead, an unknown set of test patterns. 
It does so by \textit{selecting combinations of trigger nets that are compatible but not likely to be activated simultaneously.} 
Thus, the HTs generated by~\mytool~are not likely to be activated by the logic testing-based detection techniques  (\S\ref{sec:evaluation_logic_testing}). 
On the other hand, side channel-based detection techniques apply test patterns to maximize the impact of the HT on side-channel modalities (\textit{i.e.,} switching activity).
To have a measurable impact (beyond the noise level) on switching activity, the HT needs to be activated by the defender~\cite{huang2016mers,lyu2021maxsense}. 
However, since~\mytool~generates HTs that are not likely to be activated, these HTs also evade side channel-based detection techniques, as shown in \S\ref{sec:eval_side_channel_HTs}.
Thus, the HTs generated by~\mytool~evade both logic testing- and side channel-based detection techniques considered in this work, as evidenced by our results in the next section.
\section{Results}
\label{sec:experiments}

\begin{table*}[tb]
\centering
\caption{Comparison of attack success rate (defined as 100$-$HT activation rate) for random HTs and~\mytool-generated HTs for logic testing-based detection techniques. All HTs have a trigger width ($T_{wid})$ of four. All numbers are in percentage.}
\label{tab:results_logic_testing}
\resizebox{\textwidth}{!}{
\begin{tabular}{ccccccccccccccccc}
\toprule
\multirow{3}{*}{Design} & \multirow{3}{*}{\# gates} & \multirow{1}{*}[-0.1cm]{\#}  & \multicolumn{2}{c}{Random} & \multicolumn{2}{c}{MERO~\cite{chakraborty2009mero}} & \multicolumn{2}{c}{GA+SAT~\cite{GA_SAT}} & \multicolumn{2}{c}{TGRL~\cite{pan2021automated}} & \multicolumn{3}{c}{TARMAC~\cite{TARMAC_TCAD}} & \multicolumn{3}{c}{All techniques combined}\\ 
\cmidrule(lr){4-5} 
\cmidrule(lr){6-7} 
\cmidrule(lr){8-9} \cmidrule(lr){10-11} \cmidrule(lr){12-14} \cmidrule(lr){15-17}
& & rare  & Random & RL & Random & RL & Random & RL & Random & RL & Random & RL & Improv- & Random & RL & Improv-\\
&  & nets & HTs & HTs & HTs & HTs & HTs & HTs & HTs & HTs & HTs & HTs & ement& HTs & HTs & ement
\\ \midrule
\texttt{c6288} & 2,416 & 186 & \multicolumn{1}{c}{68} & 77 & \multicolumn{1}{c}{1} & 40 & \multicolumn{1}{c}{1} & 55 & \multicolumn{1}{c}{21} & 48 & \multicolumn{1}{c}{0} & \multicolumn{1}{c}{41} & \textbf{$>$41}$\times$ & \multicolumn{1}{c}{0} & \multicolumn{1}{c}{10} & \textbf{$>$10}$\times$\\ 
\texttt{c7552} & 3,513 & 282 & \multicolumn{1}{c}{91} & 94 & \multicolumn{1}{c}{61} & 88 & \multicolumn{1}{c}{50} & 98 & \multicolumn{1}{c}{21} & 64 & \multicolumn{1}{c}{0} & \multicolumn{1}{c}{54} & \textbf{$>$54}$\times$ & \multicolumn{1}{c}{0} & \multicolumn{1}{c}{44} & \textbf{$>$44}$\times$\\ 
\texttt{s13207}  & 1,801& 604 & \multicolumn{1}{c}{97} & 100 & \multicolumn{1}{c}{85} & 99 & \multicolumn{1}{c}{21} & 96 & \multicolumn{1}{c}{97} & 99 & \multicolumn{1}{c}{1} & \multicolumn{1}{c}{71} &\textbf{71}$\times$ & \multicolumn{1}{c}{1} & \multicolumn{1}{c}{69} & \textbf{69}$\times$\\ 
\texttt{s15850}  & 2,412& 649 & \multicolumn{1}{c}{100} & 100 & \multicolumn{1}{c}{92} & 93 & \multicolumn{1}{c}{34} & 89 & \multicolumn{1}{c}{97} & 99 & \multicolumn{1}{c}{2} & \multicolumn{1}{c}{67} &\textbf{33.5}$\times$ & \multicolumn{1}{c}{2} & \multicolumn{1}{c}{57} & \textbf{28.5}$\times$ \\ 
\texttt{MIPS}  & 23,511& 1,005 & \multicolumn{1}{c}{100} & 100 & \multicolumn{1}{c}{100} & 100 & \multicolumn{1}{c}{61} & 64 & \multicolumn{1}{c}{--} & -- & \multicolumn{1}{c}{1} & \multicolumn{1}{c}{88} &\textbf{88}$\times$
& \multicolumn{1}{c}{1} & \multicolumn{1}{c}{54} & \textbf{54}$\times$\\
\texttt{AES}  & 161,161 & 854 & \multicolumn{1}{c}{100} & 100 & \multicolumn{1}{c}{100} & 100 & \multicolumn{1}{c}{100} & 100 & \multicolumn{1}{c}{--} & -- & \multicolumn{1}{c}{48} & \multicolumn{1}{c}{99} & \textbf{2.06}$\times$ & \multicolumn{1}{c}{48} & \multicolumn{1}{c}{99} & \textbf{2.06}$\times$\\
\texttt{GPS}  & 193,141 & 853& \multicolumn{1}{c}{100} & 100 & \multicolumn{1}{c}{100} & 100 & \multicolumn{1}{c}{96} & 100 & \multicolumn{1}{c}{--} & -- & \multicolumn{1}{c}{62} & \multicolumn{1}{c}{95} & \textbf{1.53}$\times$ & \multicolumn{1}{c}{62} & \multicolumn{1}{c}{95} & \textbf{1.53}$\times$\\ 
\texttt{mor1kx}  & 158,265& 641 & \multicolumn{1}{c}{100} & 100 & \multicolumn{1}{c}{99} & 100 & \multicolumn{1}{c}{49} & 100 & \multicolumn{1}{c}{--} & -- & \multicolumn{1}{c}{1} & \multicolumn{1}{c}{87} &\textbf{87}$\times$ & \multicolumn{1}{c}{1} & \multicolumn{1}{c}{87} & \textbf{87}$\times$\\ \cmidrule{1-17}
Average &  &  & \multicolumn{1}{c}{94.5} & 96.38 & \multicolumn{1}{c}{79.75} & 90 & \multicolumn{1}{c}{51.5} & 88 & \multicolumn{1}{c}{59} & 77.5 & \multicolumn{1}{c}{14.37} & \multicolumn{1}{c}{75.25} & \textbf{$>$47.26}$\times$& \multicolumn{1}{c}{14.37} & \multicolumn{1}{c}{64.37} & \textbf{$>$37.01}$\times$\\ \bottomrule
\end{tabular}
}
\end{table*}

We now explain our experimental setup and evaluate the performance of~\mytool~in evading state-of-the-art hardware Trojan (HT) detection techniques that are well-cited in academia and published in premier security venues.
We showcase the attack success rates against logic testing-based (MERO~\cite{chakraborty2009mero}, GA+SAT~\cite{GA_SAT}, TGRL~\cite{pan2021automated}, TARMAC~\cite{TARMAC_TCAD}) and side channel-based (MERS, MERS-h, MERS-s~\cite{huang2016mers}, MaxSense~\cite{lyu2021maxsense}) detection techniques.
Next, we demonstrate~\mytool's ability to generate stealthy HTs of different sizes (\textit{i.e.,} trigger widths) followed by showcasing the ramifications of HTs in performing cross-layer attacks.

\subsection{Experimental Setup}
\label{sec:setup}

\textbf{Implementation Setup.} We implement~\mytool~using \textit{PyTorch 1.12} and \textit{stable-baselines3} and train it using a Linux machine with AMD \textit{Ryzen Threadripper PRO 3975WX} with an NVIDIA \textit{A5000} GPU.
We use the Proximal Policy Optimization~\cite{schulman2017proximal} as our reinforcement learning (RL) algorithm.
We have a two layered ($64\times64$) fully-connected neural network with Tanh activation function for our policy and value networks.
We select the reward parameters, $\rho_1$ and $\rho_2$, as $-1000$ and $500$, respectively (\S\ref{sec:methodology}).
These values are chosen so that reward is positive only when the agent selects rare nets that are compatible (\textit{i.e.,} would form a valid HT), and the HT is likely to evade the test patterns.
We set $T$, the number of iterations in the offline characterization algorithm (\S\ref{sec:offline_characterization}), equal to the number of test patterns used for the logic testing-based detection techniques.
We set the parameter $S$, which controls the number of compatibility calculations and the runtime for the calculations (\S\ref{sec:trimming}),
to be $2$ since the number of compatibility calculations for $S = 3$ is much larger.
For example, for \texttt{MIPS} with $1005$ rare nets, for $S=2$, the number of calculations is $504,510$ whereas, for $S=3$, it is $\approx 1.6 \times 10^{8}$.
We set the parameter $B$, which controls the number of branches we do not prune in \S\ref{sec:pruning}, as $1$ because that provides enough HTs for the agent to explore and allows the agent to perform about $7,000$ episodes of training in $12$ hours.
Based on empirical observations, we set the threshold $\mathcal{G}$, which decides whether to employ pruning or not, as $100,000$ (\S\ref{sec:final_architecture}). We provide the hyperparameters and training reward trends in Appendix~\ref{app:additional_results_1}.

\noindent\textbf{Evaluation Setup.} We implement MERO~\cite{chakraborty2009mero}, GA+SAT~\cite{GA_SAT}, TARMAC~\cite{TARMAC_TCAD}, MERS~\cite{huang2016mers}, MERS-h~\cite{huang2016mers}, MERS-s~\cite{huang2016mers}, and MaxSense~\cite{lyu2021maxsense} in \textit{Python 3.6},
as they are not publicly available.\footnote{We reached out to the respective authors of the HT detection techniques but we did not receive the test patterns at the time of submission.}
We set a timeout of $24$ hours on all detection techniques for generating the test patterns.
We received test patterns for TGRL from the authors of~\cite{pan2021automated}. 
Additionally, we also generate random patterns for each design.
Please note that since~\mytool~is agnostic to the test patterns (\S\ref{sec:offline_characterization}), we only require test patterns from detection techniques to evaluate our generated HTs.
We generate random HTs following the procedure outlined in the detection techniques, \textit{i.e.,} we randomly selected the trigger nets from the set of rare nets.
We verified the validity of random HTs and~\mytool~generated HTs using a Boolean satisfiability check.
We evaluated $100$ HTs in both cases and set the trigger width, $T_{wid}$, (\textit{i.e.,} the number of rare nets constituting the trigger) to be $4$. 
Note that HTs with smaller trigger widths are easier to detect than HTs with larger trigger widths since the probability of activating the trigger is proportional to the product of the probabilities of rare nets being set to their rare values.
In our experiments (except \S\ref{sec:impact_different_size_HTs}), we set the trigger width as $4$ to constrain the adversary. 
Interested readers may refer to Appendix~\ref{app:appendix_experimental_flow} for the experimental flow.

\noindent\textbf{Designs.} In addition to the designs used by t\textbf{}he prior detection techniques (\textit{i.e.,} the ISCAS-85 and ISCAS-89 benchmarks widely used by the hardware security community~\cite{alkabani2007active,el2015integrated,yasin2017provably}), we also performed experiments on the \texttt{MIPS} processor from OpenCores~\cite{OpenCores_MIPS}, \texttt{AES}~\cite{aes_bomberman}, \texttt{GPS} module from \textit{Common Evaluation Platform}~\cite{gps_cep}, and \texttt{mor1kx}~\cite{mor1kx_git}. 
Since the HT detection techniques assume full-scan access,\footnote{Full-scan access enables the defender to control the values of all flip-flops in the designs,
representing the best case scenario for the defender.} for sequential designs, we modified the designs accordingly~\cite{chakraborty2009mero,GA_SAT,TARMAC_TCAD,pan2021automated,huang2016mers,lyu2021maxsense}.
Following prior work, the rareness thresholds are: $0.1$ for the ISCAS benchmarks, 9e-4 for \texttt{MIPS}, 7.2e-3 for \texttt{AES}, 4e-3 for \texttt{GPS}, and 1e-4 for \texttt{mor1lx}. 
These thresholds are chosen to have about $1,000$ rare nets to enable a fair comparison with prior work~\cite{chakraborty2009mero,GA_SAT,TARMAC_TCAD,pan2021automated,huang2016mers,lyu2021maxsense}.

\subsection{Evaluation Against Logic testing Techniques}
\label{sec:evaluation_logic_testing}

To evaluate the efficacy of~\mytool, we define the metric \textbf{attack success rate} as the percentage of HTs that evade the detection techniques. 
Mathematically, it is equivalent to $100$ $-$ HT activation rate (\S\ref{sec:prior_work}).
We compare the attack success rates of randomly generated HTs and~\mytool~generated HTs against logic testing-based detection techniques in Table~\ref{tab:results_logic_testing}.

Since we obtain the TGRL test patterns from~\cite{pan2021automated}, we use the same number of test patterns for other logic testing-based HT detection techniques
to enable a fair comparison.
However, the test patterns did not match the designs for \texttt{s13207} and \texttt{s15850}. 
Hence, the success rates of randomly generated HTs are high for these designs.
Also, since TGRL does not show results for \texttt{MIPS}, \texttt{AES}, \texttt{GPS}, and \texttt{mor1kx}~\cite{pan2021automated}, the corresponding cells in Table~\ref{tab:results_logic_testing} are empty.

\begin{table*}[tb]
\centering
\caption{Comparison of side-channel sensitivities (lower value denotes more successful attack) for random HTs and ~\mytool-generated HTs for side channel-based detection techniques. 
All HTs have a trigger width of four.
All numbers are in percentage.}
\label{tab:results_side_channel}
\resizebox{\textwidth}{!}{
\begin{tabular}{ccccccccccccccc}
\toprule
\multirow{3}{*}{Design} & \multicolumn{2}{c}{Random} & \multicolumn{2}{c}{MERS~\cite{huang2016mers}} & \multicolumn{2}{c}{MERS-h~\cite{huang2016mers}} & \multicolumn{2}{c}{MERS-s~\cite{huang2016mers}} & \multicolumn{3}{c}{MaxSense~\cite{lyu2021maxsense}} & \multicolumn{3}{c}{All techniques combined} \\
\cmidrule(lr){2-3} 
\cmidrule(lr){4-5} 
\cmidrule(lr){6-7} \cmidrule(lr){8-9} \cmidrule(lr){10-12} \cmidrule(lr){13-15}
& Random & RL & Random & RL & Random & RL & Random & RL & Random & RL & Improv- & Random & RL & Improv- \\
& HTs & HTs & HTs & HTs & HTs & HTs & HTs & HTs & HTs & HTs & ement & HTs & HTs & ement
\\ \midrule
\texttt{c6288} & \multicolumn{1}{c}{1.13} & 0.71 & \multicolumn{1}{c}{2.26} & 1.19 & \multicolumn{1}{c}{5.78} & 2.61 & \multicolumn{1}{c}{6.08} & 2.85 & \multicolumn{1}{c}{19.19} & 6.50 & \textbf{2.95$\times$} & 19.21 & 7.18 & \textbf{2.67$\times$}\\ 
\texttt{c7552} & \multicolumn{1}{c}{0.21} & 0.23 & \multicolumn{1}{c}{0.48} & 0.34 & \multicolumn{1}{c}{0.56} & 0.41 & \multicolumn{1}{c}{0.83} & 0.58 & \multicolumn{1}{c}{57.25} & 16.83 & \textbf{3.40$\times$} & 57.25 & 16.83 & \textbf{3.40$\times$} \\ 
\texttt{s13207} & \multicolumn{1}{c}{0.40} & 0.51 & \multicolumn{1}{c}{0.56} & 0.59 & \multicolumn{1}{c}{0.64} & 0.64 & \multicolumn{1}{c}{0.67} & 0.68 & \multicolumn{1}{c}{49.30} & 10.27 & \textbf{4.79$\times$} & 49.30 & 10.27 & \textbf{4.79$\times$} \\ 
\texttt{s15850} & \multicolumn{1}{c}{0.29} & 0.32 & \multicolumn{1}{c}{0.38} & 0.40 & \multicolumn{1}{c}{0.42} & 0.44 & \multicolumn{1}{c}{0.44} & 0.47 & \multicolumn{1}{c}{47.55} & 13.24 & \textbf{3.59$\times$} & 47.55 & 13.24 & \textbf{3.59$\times$} \\
\texttt{MIPS} & \multicolumn{1}{c}{0.02} & 0.02 & \multicolumn{1}{c}{0.03} & 0.03 & \multicolumn{1}{c}{0.03} & 0.03 & \multicolumn{1}{c}{0.03} & 0.02 & \multicolumn{1}{c}{2.12} & 2.48 & \textbf{0.85$\times$} & 2.12 & 2.48 & \textbf{0.85$\times$} \\
\texttt{AES} & \multicolumn{1}{c}{5.74e-3} & 4.97e-3 & \multicolumn{1}{c}{3.27e-3} & 4.09e-3 & \multicolumn{1}{c}{3.29e-3} & 4.12e-3 & \multicolumn{1}{c}{3.10e-3} & 3.65e-3 & \multicolumn{1}{c}{40.91} & 0.02 & \textbf{1667.80$\times$} & 40.91 & 0.02 & \textbf{1667.80$\times$} \\
\texttt{GPS} & \multicolumn{1}{c}{3.82e-3} & 4.08e-3 & \multicolumn{1}{c}{3.34e-3} & 2.14e-3 & \multicolumn{1}{c}{3.23e-3} & 2.10e-3 & \multicolumn{1}{c}{2.82e-3} & 2.12e-3 & \multicolumn{1}{c}{0.19} & 0.04 & \textbf{4.05$\times$} & 0.19 & 0.04 & \textbf{4.04$\times$} \\
\texttt{mor1kx} & \multicolumn{1}{c}{2.48e-3} & 2.80e-3 & \multicolumn{1}{c}{1.90e-3} & 0.91e-3 & \multicolumn{1}{c}{1.78e-3} & 0.98e-3 & \multicolumn{1}{c}{1.90e-3} & 0.91e-3 & \multicolumn{1}{c}{1.19} & 2.79 & \textbf{0.42$\times$} & 1.19 & 2.79 & \textbf{0.42$\times$} \\ \cmidrule{1-15}
Average & \multicolumn{1}{c}{0.26} & 0.22 & \multicolumn{1}{c}{0.46} & 0.32 & \multicolumn{1}{c}{0.93} & 0.51 & \multicolumn{1}{c}{1.01} & 0.57 & \multicolumn{1}{c}{27.21} & 6.52 & \textbf{210.98$\times$} & 27.21 & 6.61 & \textbf{210.94$\times$} \\ \bottomrule
\end{tabular}
}
\end{table*}

Our results demonstrate the attack success rates of~\mytool~generated HTs to be significantly higher than randomly generated HTs, \textit{i.e.,} our HTs are $\mathbf{47.26\times}$ more successful in evading the state-of-the-art logic-testing technique, TARMAC~\cite{TARMAC_TCAD}. 
On average,~\mytool~generated HTs achieve a success rate of $96.38\%$, $90\%$, $88\%$, $77.5\%$, and $75.25\%$ against test patterns generated from random, MERO~\cite{chakraborty2009mero}, GA+SAT~\cite{GA_SAT}, TGRL~\cite{pan2021automated}, and TARMAC~\cite{TARMAC_TCAD}, respectively. 
In contrast, the success rates of randomly generated HTs are $94.5\%$, $79.75\%$, $51.5\%$, $59\%$, and $14.37\%$, respectively.

Additionally, since in realistic scenarios, the defender is not constrained to rely on a single detection technique, we also evaluate the attack success rates against test patterns combined with all detection techniques considered in this work.
Even in such a scenario, our attack success rates are $\mathbf{37\times}$ higher than the success rates of randomly generated HTs.
Hence, our results corroborate that the HTs used by prior detection techniques~\cite{chakraborty2009mero,TARMAC_TCAD,GA_SAT,pan2021automated} provide a false sense of security to the designers. 
Under the assumption of randomly generated HTs, the state-of-the-art detection technique (TARMAC~\cite{TARMAC_TCAD}) detects $85\%$ of the HTs; however, against a motivated adversary (such as ours), TARMAC detects only $24\%$ of the HTs. 
\textbf{Our results highlight that we can successfully insert stealthy HTs even when a defender uses all logic testing-based detection techniques considered in this work.}

Recently, Sarihi \textit{et al.}~\cite{sarihi2022hardware} proposed a technique that utilizes RL to insert HTs.
However, there are several fundamental differences between our work and~\cite{sarihi2022hardware}, which are elucidated in \S\ref{sec:related_work_HTs}.
We obtained the HT-inserted netlists from the authors of~\cite{sarihi2022hardware} and evaluated the efficacy of their HTs through test patterns generated by TARMAC. As shown in Table~\ref{tab:GLSVLSI_results_new} the attack success rates of their HTs are close to $0\%$.\footnote{Appendix~\ref{app:additional_results_for_glsvlsi} details breakdown of success rates for HTs of different trigger widths.} 
In comparison,~\mytool-generated HTs have a success rate of $58.25\%$ for designs of similar sizes (<7K gates).

\begin{table}[tb]
\centering
\caption{Attack success rate (\%) for HTs generated by~\cite{sarihi2022hardware} against TARMAC~\cite{TARMAC_TCAD}.}
\label{tab:GLSVLSI_results_new}
\resizebox{0.46\textwidth}{!}{
\begin{tabular}{cccccccc}
\toprule
\rotatebox{0}{Design} & \rotatebox{0}{\texttt{c432}} & \rotatebox{0}{\texttt{c880}} & \rotatebox{0}{\texttt{c1355}} & \rotatebox{0}{\texttt{c1908}} & \rotatebox{0}{\texttt{c3540}} & \rotatebox{0}{\texttt{c6288}} & \rotatebox{0}{Average} \\
\midrule
Success Rate & 0 & 0 & 0.02 & 0.15 & 0 & 0.02 & 0.03 \\
\bottomrule
\end{tabular}
}
\end{table}

\subsection{Evaluation Against Side-channel Techniques}
\label{sec:eval_side_channel_HTs}

\begin{table}[tb]
\centering
\caption{Comparison of percentage of HTs with side-channel sensitivity less than 10\%, for randomly generated and~\mytool~generated HTs against MaxSense~\cite{lyu2021maxsense}.}
\label{tab:MaxSense_detection_rate}
\resizebox{0.46\textwidth}{!}{
\begin{tabular}{cccccccccc}
\toprule
\rotatebox{45}{Design} & \rotatebox{45}{\texttt{c6288}} & \rotatebox{45}{\texttt{c7552}} & \rotatebox{45}{\texttt{s13207}} & \rotatebox{45}{\texttt{s15850}} & \rotatebox{45}{\texttt{MIPS}} & \rotatebox{45}{\texttt{AES}} & \rotatebox{45}{\texttt{GPS}} & \rotatebox{45}{\texttt{mor1kx}} & \rotatebox{45}{Average} \\
\midrule
Random HTs & 27 & 19 & 5 & 13 & 100 & 0 & 100 & 99 & 45.37\\
\midrule
RL HTs & 81 & 71 & 60 & 60 & 97 & 100 & 100 & 97 & 83.25\\
\midrule
Improv. & 3.00$\times$ & 3.73$\times$ & 12$\times$ & 4.61$\times$ & 0.97$\times$ & $>$100$\times$ & 1$\times$ & 0.97$\times$ & \textbf{$>$15.78$\times$} \\
\bottomrule
\end{tabular}
}
\end{table}

We evaluate the success rates of HTs generated by~\mytool~in evading the side channel-based detection techniques in Table~\ref{tab:results_side_channel}.
To that end, we compare the side-channel sensitivity (a metric chosen by prior~\cite{huang2016mers} and state-of-the-art side channel-based detection techniques~\cite{lyu2021maxsense}) of randomly generated HTs with~\mytool-generated HTs in Table~\ref{tab:results_side_channel}.
Recall that side-channel sensitivity measures the amount of switching caused by an HT relative to the switching in an HT-free design (\S\ref{sec:background}). 
A lower sensitivity value implies that the impact of an HT on the side-channel metrics is overshadowed by the impact of the regular circuit activity and environmental noise. 
Therefore, lower values of side-channel sensitivity indicate a more successful attack.
On average, the sensitivity of test patterns generated from MaxSense~\cite{lyu2021maxsense} (the state-of-the-art side channel-based technique) for~\mytool-generated HTs is $210.98\times$ lower than randomly generated HTs.
We also perform an experiment to combine the test patterns generated from all detection techniques, in such a scenario, the sensitivity of the combined set of test patterns for~\mytool-generated HTs is $210.94\times$ lower than randomly generated HTs.
As Table~\ref{tab:results_side_channel} and~\cite{lyu2021maxsense} demonstrate, MaxSense is by far the best side channel-based HT detection technique (among the techniques considered in this work), so, in the remainder of this section, we perform further analysis of its efficacy. 
In particular, we analyze the percentage of HTs evading detection from side channel-based detection techniques based on the threshold (side-channel sensitivity) of $10\%$ used by prior works~\cite{balaji2012accurate_sensitivity_threshold,lyu2021maxsense}.
An HT is considered detected if the sensitivity is greater than the threshold, \textit{i.e.,} $10\%$.
Table~\ref{tab:MaxSense_detection_rate} demonstrates the percentage of HTs evading detection from MaxSense~\cite{lyu2021maxsense} (results for other techniques are shown in Appendix~\ref{app:additional_results_2}). 
On average, $83\%$ of~\mytool~generated HTs evade detection from MaxSense~\cite{lyu2021maxsense}, whereas, for randomly generated HTs (\textit{i.e.,} the evaluation approach used by prior work), only $45\%$ of the HTs evade detection.
\textbf{Our results highlight that we can successfully insert HTs even when a defender uses all side channel-based detection techniques considered in this work.}

\begin{table}[tb]
\centering
\caption{Comparison of sensitivity to power consumption for randomly generated and~\mytool~generated HTs against MaxSense~\cite{lyu2021maxsense}.}
\label{tab:results_power_consumption}
\resizebox{0.46\textwidth}{!}{
\begin{tabular}{cccccccccc}
\toprule
\rotatebox{45}{Design} & \rotatebox{45}{\texttt{c6288}} & \rotatebox{45}{\texttt{c7552}} & \rotatebox{45}{\texttt{s13207}} & \rotatebox{45}{\texttt{s15850}} & \rotatebox{45}{\texttt{MIPS}} & \rotatebox{45}{\texttt{AES}} & \rotatebox{45}{\texttt{GPS}} & \rotatebox{45}{\texttt{mor1kx}} & \rotatebox{45}{Average} \\
\midrule
Random HTs & 18.84 & 56.32 & 52.41 & 53.89  & 0.23 & 0.18 & 0.02 & 3.83 & 23.21\\
\midrule
RL HTs & 6.68 & 15.97 & 11.23 & 16.38 & 0.06 & 0.03 & 0.02 & 0.28 & 6.33\\
\midrule
Improv. & &  & & & & & &  &\\
in stealthiness & \multirow{-2}{*}{2.81$\times$}  & \multirow{-2}{*}{3.52$\times$} & \multirow{-2}{*}{4.66$\times$} & \multirow{-2}{*}{3.28$\times$} & \multirow{-2}{*}{3.55$\times$} &  \multirow{-2}{*}{5.14$\times$} & \multirow{-2}{*}{$\sim1\times$} & \multirow{-2}{*}{13.53$\times$}  & \multirow{-2}{*}{\textbf{4.68$\times$}}\\
\bottomrule
\end{tabular}
}
\end{table}

The side channel-based HT detection techniques considered in this work evaluate the efficacy using logic simulation-based sensitivity analysis as we did in Table~\ref{tab:results_side_channel}.
To further validate the efficacy of~\mytool, we performed power simulations to compare the sensitivity of~\mytool-generated HTs and randomly generated HTs against MaxSense test patterns. 
In particular, following the user guides of industrial tools, we executed the industrial tool flow to measure the stealth of~\mytool-generated HTs in evading power-based side-channel analysis. 
Given the HT-inserted netlists and MaxSense test patterns, we perform logic simulations using Synopsys VCS to obtain a value change dump (VCD) that contains information about the switching activity of all nets in the netlists. 
This VCD and the academic \textit{NanGateOpenCell45nm} library files are provided to Synopsys PrimeTime (a widely-used industrial tool used by semiconductor companies for power simulations to accurately predict the power characteristics of a manufactured chip) to obtain power consumption traces for the MaxSense test patterns.

The power traces for the randomly generated HTs and~\mytool-generated HTs are compared with the power trace for the HT-free netlist to calculate the percentage deviations in the power consumption, \textit{i.e.,} sensitivity to power consumption, for the randomly- and~\mytool-generated HTs.
Table~\ref{tab:results_power_consumption} depicts the sensitivity to power consumption for the randomly- and~\mytool-generated HTs. 
On average, the sensitivity of~\mytool-generated HTs is $4.68\times$ lower than randomly-generated HTs.

\begin{figure}[tb]
\centering
\includegraphics[width=0.475\textwidth,trim=0cm 0cm 0cm 1.1cm, clip]{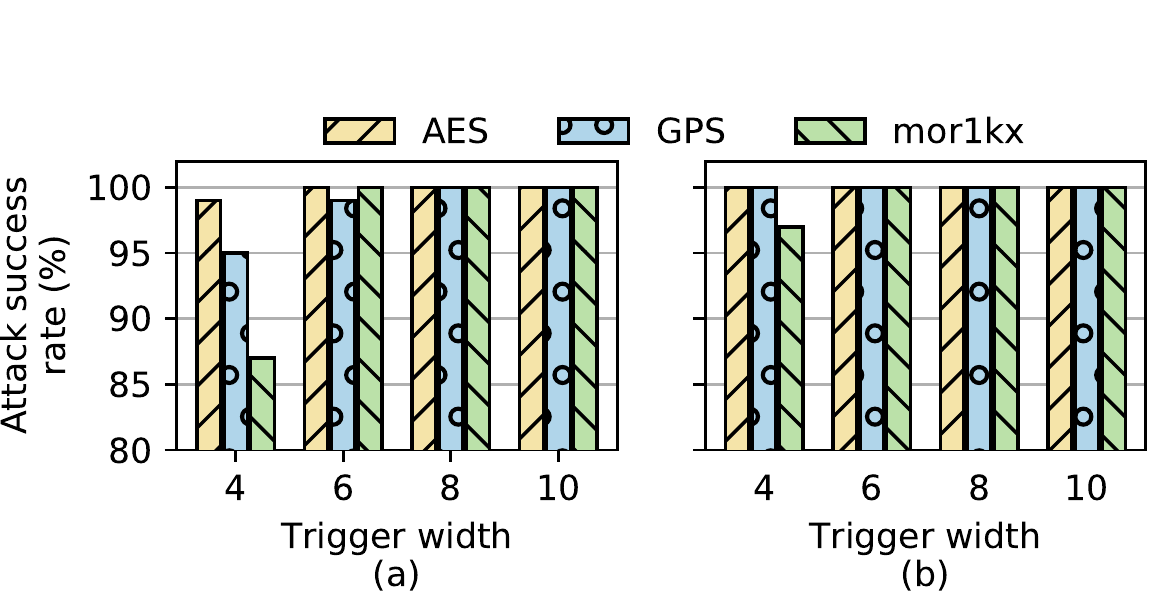}
\smallerspacecaption
\smallerspacecaption
\smallerspacecaption
\caption{Impact of HT size (\textit{i.e.,} trigger width) on attack success rates against (a)~TARMAC~\cite{TARMAC_TCAD} and (b)~MaxSense~\cite{lyu2021maxsense} for \texttt{AES}, \texttt{GPS}, and \texttt{mor1kx}.}
\label{fig:trig_width_analysis_against_TARMAC_MaxSense_aes_gps_mor1kx}
\end{figure}

\subsection{Impact of Different Sizes of HTs}
\label{sec:impact_different_size_HTs}

Now, we analyze the efficacy of~\mytool~generated HTs as trigger width ($T_{wid}$) increases.
We generate HTs for the largest designs (\texttt{AES}, \texttt{GPS}, and \texttt{mor1kx}) for trigger widths of $4$, $6$, $8$, and $10$.
Figure~\ref{fig:trig_width_analysis_against_TARMAC_MaxSense_aes_gps_mor1kx} (a) illustrates the attack success rates against TARMAC~\cite{TARMAC_TCAD}, the state-of-the-art technique in logic testing-based detection.\footnote{We do not perform experiments for MERO~\cite{chakraborty2009mero} and GA+SAT~\cite{GA_SAT} as our attack success rate against these techniques is already $100\%$ for a trigger width of $4$.}
As the trigger width increases,~\mytool~generated HTs evade TARMAC~\cite{TARMAC_TCAD} (100\% attack success rate) for all the designs.

We observe similar results against MaxSense (Figure~\ref{fig:trig_width_analysis_against_TARMAC_MaxSense_aes_gps_mor1kx}(b)). 
As the trigger width increases, the attack success rate increases. 
At first glance, this result might seem counter-intuitive since an HT with larger trigger width requires more logic gates, which means that the impact of the HT on switching would be larger. 
However, since our HTs are stealthy, they are not activated by the test patterns and hence have no impact on switching, leading to a success rate of 100\% for larger trigger widths.

\subsection{Is Reinforcement Learning Necessary?}
\label{sec:compare_attrition}

Till now, our results demonstrated that ~\mytool~(consisting of domain-specific optimizations and RL agent) generates stealthy HTs that evade the detection techniques considered in this work.
Here, we showcase that RL is an integral part of~\mytool~and crucial to generating stealthy HTs.
To that end, we develop a new HT-insertion algorithm called ``\mytool~minus RL.''
In other words, we use the same compatibility information, trimming, and detection metrics as~\mytool, but generate HTs \textit{without} using an RL agent.
In particular, we generate HTs iteratively.
In each iteration, a set of trigger nets are chosen and evaluated using the characterization, compatibility information, and trimming outlined in \S\ref{sec:offline_characterization} and \S\ref{sec:trimming}.
If trigger nets are (i)~not compatible or (ii)~activated by any of the sets ($\mathcal{C}_i$) from offline characterization, they are discarded.
Otherwise, we generate an HT.
To enable fair comparison, this algorithm is executed for the same duration as the RL training.
We compare the success rates of~\mytool~ minus RL and~\mytool~against TARMAC in Table~\ref{tab:attrition_minus_RL_and_tetramax_HTs_comparison_copy}.

\mytool~minus RL achieves an average success rate of $49.62\%$ (Table~\ref{tab:attrition_minus_RL_and_tetramax_HTs_comparison_copy}) whereas~\mytool~has a success rate of $75.25\%$ against TARMAC (Table~\ref{tab:results_logic_testing}). 
Moreover~\mytool~has a success rate of $1.71\times$ that of~\mytool~minus RL, showcasing that utilizing domain-specific optimizations are not sufficient and that RL plays a significant role in~\mytool~toward generating stealthy HTs.

Till now, we evaluated the efficacy of~\mytool~with regards to randomly generated HTs, which is consistent with the assumptions made in detection techniques considered in this work.
To further demonstrate the improvement in stealthiness offered by~\mytool, we consider a better approach for inserting HTs, \textit{i.e.,} we generate HTs that are guaranteed to evade test patterns generated by a commercial IC testing tool.
To that end, we generate test patterns using Synopsys TestMAX (for a given design) and use a random sampling approach to create a corpus of HTs that evade these test patterns~\cite{TestMAX}.\footnote{Since the generated HTs are guaranteed to evade test patterns; this approach is superior to randomly-generated HTs.}
We showcase the success rates of the HTs generated using TestMAX test patterns in Table~\ref{tab:attrition_minus_RL_and_tetramax_HTs_comparison_copy}. Results demonstrate the superiority of~\mytool-generated HTs; overall, we achieve a success rate of $>37.39\times$ compared to HTs generated to evade TestMAX test patterns.

\begin{table}[tb]
\centering
\caption{Comparison of attack success rate against TARMAC~\cite{TARMAC_TCAD} for ~\mytool~minus RL, Synopsys TestMAX~\cite{TestMAX}, and~\mytool.}
\label{tab:attrition_minus_RL_and_tetramax_HTs_comparison_copy}
\resizebox{0.46\textwidth}{!}{
\begin{tabular}{cccccccccc}
\toprule
\rotatebox{45}{Design} & \rotatebox{45}{\texttt{c6288}} & \rotatebox{45}{\texttt{c7552}} & \rotatebox{45}{\texttt{s13207}} & \rotatebox{45}{\texttt{s15850}} & \rotatebox{45}{\texttt{MIPS}} & \rotatebox{45}{\texttt{AES}} & \rotatebox{45}{\texttt{GPS}} & \rotatebox{45}{\texttt{mor1kx}} & \rotatebox{45}{Average} \\
\midrule
~\mytool~ & &  & & & & & &  &\\
minus RL & \multirow{-2}{*}{51}  & \multirow{-2}{*}{33} & \multirow{-2}{*}{59} & \multirow{-2}{*}{37} & \multirow{-2}{*}{25} &  \multirow{-2}{*}{72} & \multirow{-2}{*}{80} & \multirow{-2}{*}{40}  & \multirow{-2}{*}{49.62}\\
\midrule
Impr./~\mytool~ & &  & & & & & &  &\\
minus RL & \multirow{-2}{*}{0.80$\times$}  & \multirow{-2}{*}{1.63$\times$} & \multirow{-2}{*}{1.20$\times$} & \multirow{-2}{*}{1.81$\times$} & \multirow{-2}{*}{3.52$\times$} &  \multirow{-2}{*}{1.37$\times$} & \multirow{-2}{*}{1.18$\times$} & \multirow{-2}{*}{2.71$\times$}  & \multirow{-2}{*}{\textbf{1.71$\times$}}\\
\midrule
Synopsys & &  & & & & & &  &\\
TestMAX & \multirow{-2}{*}{0}  & \multirow{-2}{*}{0} & \multirow{-2}{*}{2} & \multirow{-2}{*}{2} & \multirow{-2}{*}{2} &  \multirow{-2}{*}{39} & \multirow{-2}{*}{59} & \multirow{-2}{*}{1}  & \multirow{-2}{*}{13.12}\\
\midrule
Impr./Synopsys & &  & & & & & &  &\\
TestMAX & \multirow{-2}{*}{$>$41$\times$}  & \multirow{-2}{*}{$>$54$\times$} & \multirow{-2}{*}{35.5$\times$} & \multirow{-2}{*}{33.5$\times$} & \multirow{-2}{*}{44$\times$} &  \multirow{-2}{*}{2.53$\times$} & \multirow{-2}{*}{1.61$\times$} & \multirow{-2}{*}{87$\times$}  & \multirow{-2}{*}{\textbf{$>$37.39$\times$}}\\
\bottomrule
\end{tabular}
}
\end{table}

\subsection{Ramifications of Stealthy Hardware Trojans}
\label{sec:case_studies}

Till now, our results demonstrated that~\mytool~generates HTs that evade the considered detection techniques; however, this does not necessarily 
imply that they can be used to launch practical attacks. 
Here, we demonstrate the usefulness of~\mytool~generated HTs by showcasing two case-studies 
that undermine the security of the largest processor considered in this work, \textit{i.e.,} \texttt{mor1kx}.

\begin{figure}[tb]
\centering
\includegraphics[trim=1cm 0.5cm 0.5cm 0.5cm, clip,width=0.475\textwidth]{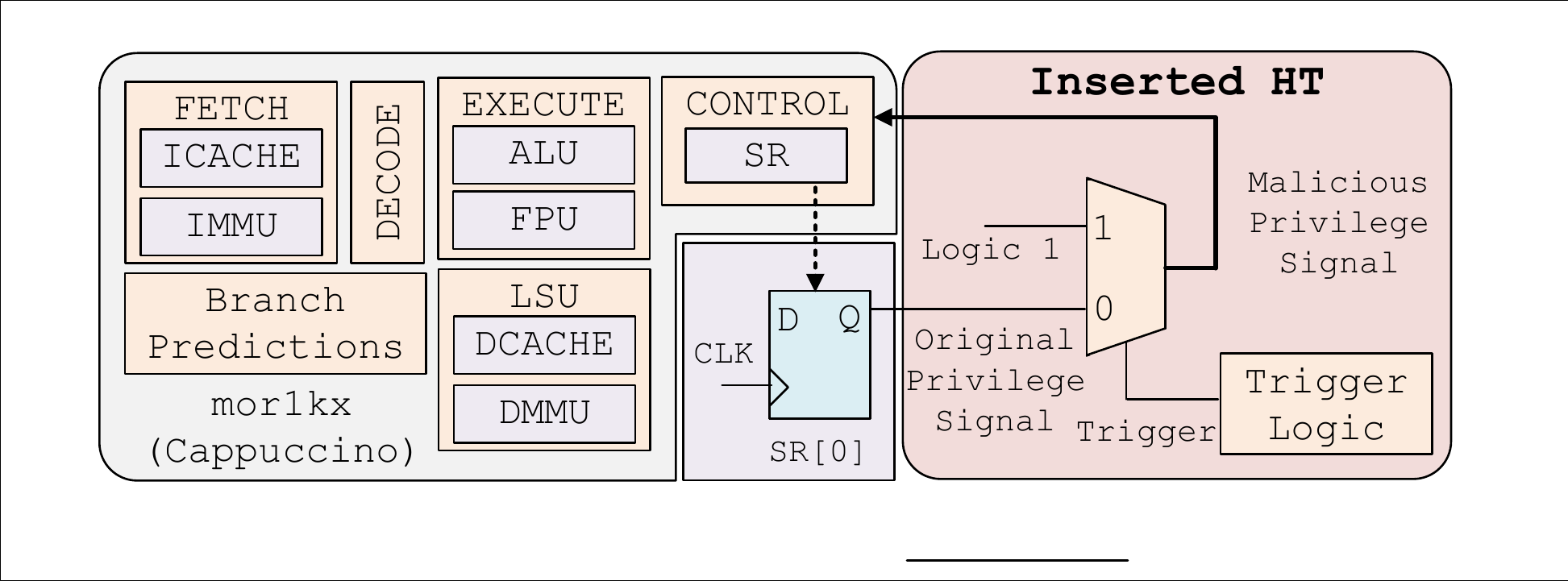}
\smallerspacecaption
\smallerspacecaption
\smallerspacecaption
\smallerspacecaption
\caption{HT performing privilege escalation in \texttt{mor1kx}.}
\label{fig:case_study}
\end{figure}

\noindent\textbf{Privilege Escalation.} To perform privilege escalation, we need to identify the high-level instructions an adversary should use to trigger the rare nets, \textit{i.e.,} the Boolean values of the rare nets need to be mapped to valid assembly-level instructions. 
Given~\mytool~generated rare nets, we write formal properties corresponding to the values of those rare nets and use \textit{Cadence JasperGold} to provide binaries that would trigger the HT. 
Finally, we translate the binaries into valid instructions using the processor architecture manual~\cite{OR1200_arch_manual}. 
Note that the devised HT that causes privilege escalation evades all detection techniques considered in this work, as confirmed by our experiments.

For \texttt{mor1kx}, one rare net chosen by~\mytool~is from the processor control unit, another is from the system bus, and the other two are from the fetch stage.
The target flip-flop to perform privilege escalation is a part of the special-purpose Supervision Register, SR. 
The least significant bit of SR holds the mode of the processor: $0$ indicates user mode and $1$ indicates supervisor mode.
Table~\ref{tab:case_studies} denotes the instruction sequence that needs to be executed to perform privilege escalation.
The attack requires $12$ clock cycles since the first instruction \texttt{l.j 0xc0000} to activate the HT that flips the privilege bit (Figure~\ref{fig:case_study}).
We illustrate the waveform of the attack execution in Appendix~\ref{app:additional_results_3}.

\begin{table}[tb]
\centering
\caption{Instruction sequences to launch a privilege escalation attack and to activate a kill switch (\textit{i.e.,} force the processor in an infinite loop) using~\mytool~generated HTs for \texttt{mor1kx}.}
\label{tab:case_studies}
\resizebox{0.48\textwidth}{!}{
\begin{tabular}{cccc}
\cmidrule(lr){1-2} 
\cmidrule(lr){3-4} 
\footnotesize{Clock Cycle} & \footnotesize{Privilege Escalation} & \footnotesize{Clock Cycle} & \footnotesize{Kill Switch}\\
\cmidrule(lr){1-2} 
\cmidrule(lr){3-4} 
\begin{tabular}[c]{@{}c@{}}\texttt{\footnotesize{1}}\\ \texttt{\footnotesize{3}}\\ \texttt{\footnotesize{6}}\\ \texttt{\footnotesize{9}}\\ \texttt{\footnotesize{12}}\end{tabular} &
\begin{tabular}[c]{@{}c@{}}\texttt{\footnotesize{l.j 0xc0000}} \\ \texttt{\footnotesize{l.bf 0x3ee7ef4}}\\ \texttt{\footnotesize{l.mtspr r7, r7, 0x0}}\\ \texttt{\footnotesize{l.j -0x123fff0}}\\ \texttt{\footnotesize{l.mulu r15, r15, r15}}\end{tabular} &
\begin{tabular}[c]{@{}c@{}}\texttt{\footnotesize{1}}\\ \texttt{\footnotesize{3}}\\ \texttt{\footnotesize{6}}\\ \texttt{\footnotesize{9}}\\
\texttt{\footnotesize{12}}\\ \texttt{\footnotesize{15}}\\\texttt{\footnotesize{18}}\end{tabular} &
\begin{tabular}[c]{@{}c@{}}\texttt{\footnotesize{l.j -0x4}}\\ \texttt{\footnotesize{l.mtspr r18, r18, 0x0}}\\ \texttt{\footnotesize{l.mtspr r9, r18, 0x1000}}\\ \texttt{\footnotesize{l.jr r2}}\\ \texttt{\footnotesize{l.mtspr r5, r25, 0x4000}}\\ \texttt{\footnotesize{l.mtspr r11, r24, 0x0}}\\ \texttt{\footnotesize{l.mulu r31, r0, r0}}\end{tabular}
 \\
\cmidrule(lr){1-2}
\cmidrule(lr){3-4} 
\end{tabular}
}
\end{table}

\noindent\textbf{Kill switch.} Here, we demonstrate a case study on how~\mytool~generated HTs can be used to
force the \texttt{mor1kx} processor in an infinite loop, \textit{i.e.,} stop further useful execution when the malicious instructions are executed.
\mytool~picks the four rare nets as follows: one from the system bus, one from the fetch stage, and two from the processor control unit. 
The target register is the program counter.
Table~\ref{tab:case_studies} denotes the instruction sequence that need to be executed to launch the attack. 
This attack requires $20$ clock cycles since the first instruction \texttt{l.j -0x4}.
Upon HT activation, the program counter gets stuck at \texttt{0x00000000} preventing further execution until a restart is performed.
Essentially, this HT is a demonstration of a ``\textit{kill switch}'' that can be activated by executing a handful of malicious instructions.

Thus, these case studies demonstrate that an adversary can repurpose~\mytool~to design HTs that not only evade detection, but also cause practical, cross-layer, and real-world attacks.
\section{Related Work}
\label{sec:related_work}

While this work assumes a hardware Trojan (HT) inserted in an untrusted foundry (the threat model highlighted by the U.S. Department of Defense~\cite{DoD_supply_chain}), we highlight how our work relates to HTs inserted during design time. 
We demonstrate an additional advantage of~\mytool~in generating \textit{dynamic} HTs as opposed to the \textit{static} HTs available in TrustHub~\cite{TrustHub}, a popular database used by the hardware security community.
Finally, we discuss how RL has been used for benchmarking defenses.

\subsection{Hardware Trojan Insertion and Evasion}
\label{sec:related_work_HTs}

HTs can be inserted during design time (adversary is the third-party intellectual property provider) or during fabrication time (adversary in the foundry).
Next, we highlight some examples of HT insertion in an untrusted foundry.
Lin \textit{et al.}~\cite{lin2009trojan} designed an HT that induced physical side-channels to leak the keys of an AES cryptographic accelerator.
Becker \textit{et al.}~\cite{becker2013stealthy} modified the dopant polarity of transistors that forces the input of the logic gates to a constant logic level of $0$ or $1$.
Yang \textit{et al.}~\cite{yang2016a2} performed privilege escalation using a capacitor-based HT on fabricated chips.

For HTs inserted during design time, researchers proposed various insertion and corresponding evasion techniques, leading to a \textit{cat-and-mouse} game, which we briefly outline next.
~\cite{hicks2010overcoming} identifies unused portions of the circuit during design-time testing and flags them as potentially malicious.
They attempt to detect HTs by identifying pairs of dependent signals in the design that could ostensibly be replaced by a wire without impacting the outcome of verification test cases.
However, Sturton \textit{et al.}~\cite{sturton2011defeating} thwarted the detection technique proposed in~\cite{hicks2010overcoming} by developing HTs such that no pair of dependent signals are always equal during design-time testing. 
Functional analysis for nearly-unused circuit identification (FANCI) identifies input wires that can serve as backdoor triggers by measuring the degree of control an input has on the outputs of a design~\cite{waksman2013fanci}. 
In addition, VeriTrust examines verification corners to identify trigger signals~\cite{zhang2015veritrust}. 
However, DeTrust~\cite{zhang2014detrust} evades FANCI~\cite{waksman2013fanci}
by spreading the trigger logic across multiple combinational logic blocks, while HT triggers are hidden into multiple sequential levels to deter VeriTrust~\cite{zhang2014detrust}.

In contrast,~\mytool: (i)~is an automated framework to generate stealthy HTs, requiring no manual intervention; (ii)~can adapt to new detection techniques and generate HTs dynamically, so, if a new detection technique is developed in the future, our framework would still hold even if the current HTs we generate may or may not;
(iii)~is not a point solution to defeat a specific detection technique unlike the efforts in the design-time attack model, so, is not subject to a cat-and-mouse game of attacks and defenses.

More recently, Sarihi \textit{et al.}~\cite{sarihi2022hardware} also utilize RL to insert HTs. 
However, several key differences exist. 
First, the modeling of states, actions, and rewards of both techniques differ considerably. 
Second,~\mytool~focuses on inserting HTs that evade detection techniques, whereas~\cite{sarihi2022hardware} insert HTs that maximize the number of inputs (\textit{i.e.,} trigger width).
Third,~\cite{sarihi2022hardware} does not evaluate the efficacy of generated HTs in evading detection techniques. 
In contrast, the prime objective of~\mytool~is generating HTs that evade detection techniques (across logic testing-based and side channel-based approaches). 
Fourth,~\mytool~incorporates three solutions using domain knowledge, rendering it scalable; such solutions providing experimental evidence of scalability are missing in~\cite{sarihi2022hardware}.
Finally,~\mytool~demonstrates results on designs up to >190K gates, whereas the largest design in~\cite{sarihi2022hardware} has only $\approx$7K gates.

\subsection{\textit{``Static''} Hardware Trojan Benchmarks}
\label{sec:prior_HT_benchmark_creation}

TrustHub~\cite{TrustHub,salmani2013design} is a database of different HT benchmarks (a total of $91$) and provides a common platform for researchers to evaluate detection techniques.
However, the HTs in TrustHub are created from randomly selected rare nets, \textit{i.e.,} they do not reflect a real-world adversary because the adversary inserts HTs in a directed manner with the objective of evading detection.
Moreover, these benchmarks were last updated in 2017, and since then, there have been several new HT detection techniques. So, the benchmarks do not keep up with the developments in HT detection since they are static, unlike our framework, which can generate HTs targeted towards new detection techniques too.

After TrustHub, there have been several works that have developed automated tools for HT insertion~\cite{shakya2017benchmarking,cruz2018automated,yu2019improved}.
However, most automated tools for HT insertion rely on signal probability, \textit{i.e.,} the generated HTs have a very low probability of getting triggered during logic testing~\cite{cruz2018automated}.
Another limitation of these tools is that they 
insert HTs in the design by randomly picking the rare nets from a pool of rare nets~\cite{cruz2018automated,yu2019improved}. 
However, an adversary is not constrained to design the trigger randomly.
Instead, an adversary would characterize the design space of the underlying design and insert HTs that are likely to evade many detection techniques.
Additionally, the benchmarks from the aforementioned HT insertion tools are ``static'' in nature, \textit{i.e.,} they have not evolved with time and new detection techniques.

In contrast,~\mytool~generates HTs in an automated manner and can readily adapt to new detection techniques, thereby generating HTs dynamically and providing a litmus test for existing (and future) HT detection techniques.
We envision that such a set of HT-infested benchmarks
can aid researchers in developing
detection techniques that do not provide a ``\textit{false sense of security}.''

\subsection{Potential Countermeasures}
\label{sec:countermeasures}

\mytool~generates stealthy HTs that evade eight detection techniques spanning logic testing and side channel-based approaches (\S\ref{sec:experiments}).
As a result, our work highlights the requirement of developing robust HT detection techniques.
Next, we outline two directions that could potentially mitigate the attack presented in this work.

The first direction involves modifying the GDSII of the design to prevent HT insertion. 
Designers can modify the GDSII by making the design congested~\cite{ICAS,knechtel2022benchmarking,cocchi2014circuit} or increasing the utilization~\cite{ICAS,knechtel2022benchmarking,cocchi2014circuit}.
Such an approach makes the targeted insertion of HTs difficult for an adversary. 
Recall that HT insertion entails augmenting logic gates and wires into the underlying design, and if there is a dearth of space (in terms of placement sites and/or routing tracks) in the GDSII, an attacker faces considerable challenges.
ICAS took the first step toward this direction~\cite{ICAS}, and the tool-chain enables designers to evaluate the susceptibility of a GDSII toward HT insertion. 
However, increasing congestion and utilization of the chip is usually impractical since timing closure becomes challenging~\cite{knechtel2022benchmarking}.
Real-world designs such as processors are rarely fabricated with utilization greater than $70\%$ to allow engineering change order-related fixes. 
This provides enough space for inserting HTs, thereby limiting the efficacy of such defenses.

Another direction is a consequence of our results in Table~\ref{tab:results_logic_testing}.
Consider the design \texttt{c6288} for which~\mytool's success rate is low. 
As analyzed and explained in Appendix~\ref{app:additional_results_1}, this is due to the structure of the design where several rare nets are compatible with each other, which increases the ease of activating many rare nets simultaneously, leading to easy detection of HTs. 
As part of future work, we intend to explore a way to mitigate attacks such as~\mytool~by making changes to the structure of a given design (through synthesis) so that many rare nets are compatible with each other. 
This way, we can make detecting HTs easier for existing detection tools.
\section{Conclusion and Ramifications}
\label{sec:conclusion}

Hardware Trojans (HTs) inserted during fabrication pose a potent threat to the security of integrated circuits.
Unfortunately, we note that state-of-the-art HT detection techniques have been tested only with \textit{weak} adversarial assumptions (\textit{i.e.,} random insertion of HTs), providing a \textit{``false sense of security''} for over a decade.
This calls for a critical rethinking of security evaluation methodologies for HT detection techniques.

Aided by the power of reinforcement learning (RL), we developed~\mytool~that automatically generates stealthy HTs that evade eight HT detection techniques from two different approaches, including the state-of-the-art. 
To that end, we overcame five challenges related to computational complexity, efficiency, scalability, and practicality using three solutions.
We showcased the efficacy of~\mytool~on different designs ranging from widely-used academic benchmarks to processors.
~\mytool~achieved average attack success rates $47\times$ and $211\times$ compared to randomly inserted HTs against the state-of-the-art logic testing and side channel techniques.
Additionally, we illustrated cross-layer attacks through two case studies (privilege escalation and kill switch) on the open-source \texttt{mor1kx} processor, demonstrating an end-to-end methodology for generating HTs that are not only stealthy, but also useful for realistic attacks.
As part of future work, we intend to examine how an RL-based defender~\cite{gohilDAC22} can spar with an RL-based attacker (such as~\mytool) to generate interesting attacks and defenses.

\noindent\textbf{Ramifications.}~\mytool~calls for action on re-evaluating the assumptions made by the HT detection techniques. Rather than considering randomly inserted HTs, developers of future HT detection techniques must consider the existence of a motivated adversary.
To that end,~\mytool~(i)~demonstrates the reduction in the efficacy of HT detection techniques when~\mytool~generated HTs are considered, (ii)~aids in creating a suite of HT-infested designs that can be used by researchers to benchmark the strength of their detection techniques, and (iii)~urges the community to develop efficient detection techniques that are scalable to real-world designs.

\section*{Acknowledgments}
The work was partially supported by the National Science Foundation (NSF CNS--1822848 and NSF DGE--2039610). 
We thank Amin~Sarihi and Prof.\ Abdel-Hameed A.\ Badawy from New Mexico State University for providing the HT designs of their work.
In addition, we thank Prof.\ Dileep~Kalathil, Zaiyan~Xu, Rahul~Kande, Chen~Chen, and the anonymous reviewers for their constructive comments.

\clearpage
\bibliographystyle{unsrt}
\bibliography{main}

\clearpage
\appendix
\section{Appendix}
\label{sec:appendix}

\subsection{Integrated Circuit Design Flow}
\label{app:appendix_supply_chain}

\begin{figure*}[tb]
\centering
\includegraphics[trim=0.2cm 0.2cm 0.2cm 0.2cm, clip, width=\textwidth]{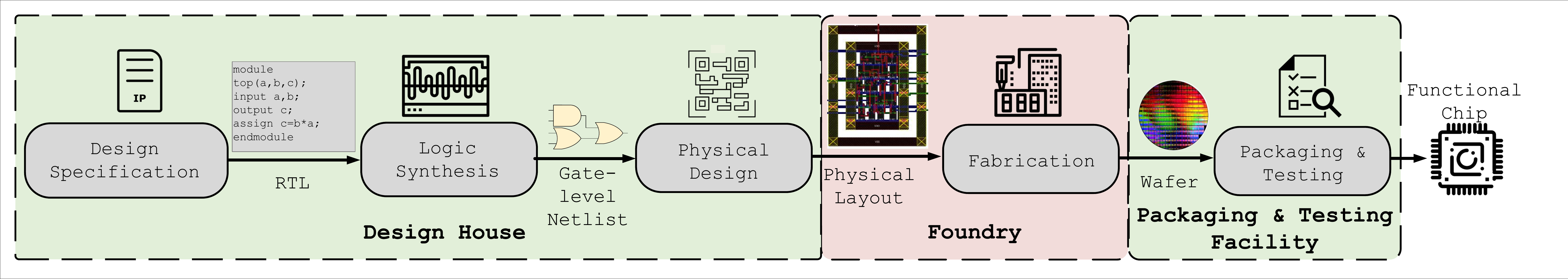}
\smallerspacecaption
\smallerspacecaption
\smallerspacecaption
\caption{Integrated circuit design flow and supply chain. 
In our work, we assume a trusted design house and trusted packaging and testing facility. Fabrication facility (a.k.a. foundry) is untrusted.}
\label{fig:IC_supply_chain}
\end{figure*}

Developing a complex integrated circuit (IC) involves several design stages heavily assisted by computer-aided design (CAD) tools.
First, the high-level design specifications are decided based on the functionality of the intended chip, and designers (or product architects) develop the optimal architecture using the hardware description language (HDL) of choice (e.g., Verilog) at the register-transfer level (RTL).
At times, the design team also deploys third-party intellectual property (IP) cores in the design.
Next, CAD tools synthesize the HDL codes into a gate-level netlist for a given technology node.
After synthesis, the design is taken through floorplanning, placement, and routing through a physical synthesis tool (commonly known as layout generation tools) to create a physical layout.
The physical layout is encoded into a graphics database system format which is then sent to a fabrication facility (a.k.a foundry) for fabrication.
Once the design is fabricated, it is subject to functional and performance testing. 
Once the fabricated chips pass the intended tests, they are packaged and sold to end-users.
We illustrate the IC design flow and the supply chain in Figure~\ref{fig:IC_supply_chain}.
Our work considers the design house and the testing and packaging facility to be trustworthy while the foundry is untrustworthy.

\begin{figure}[tb]
\centering
\includegraphics[trim=0.6cm 0.6cm 0.6cm 0.6cm, clip, width=0.475\textwidth]{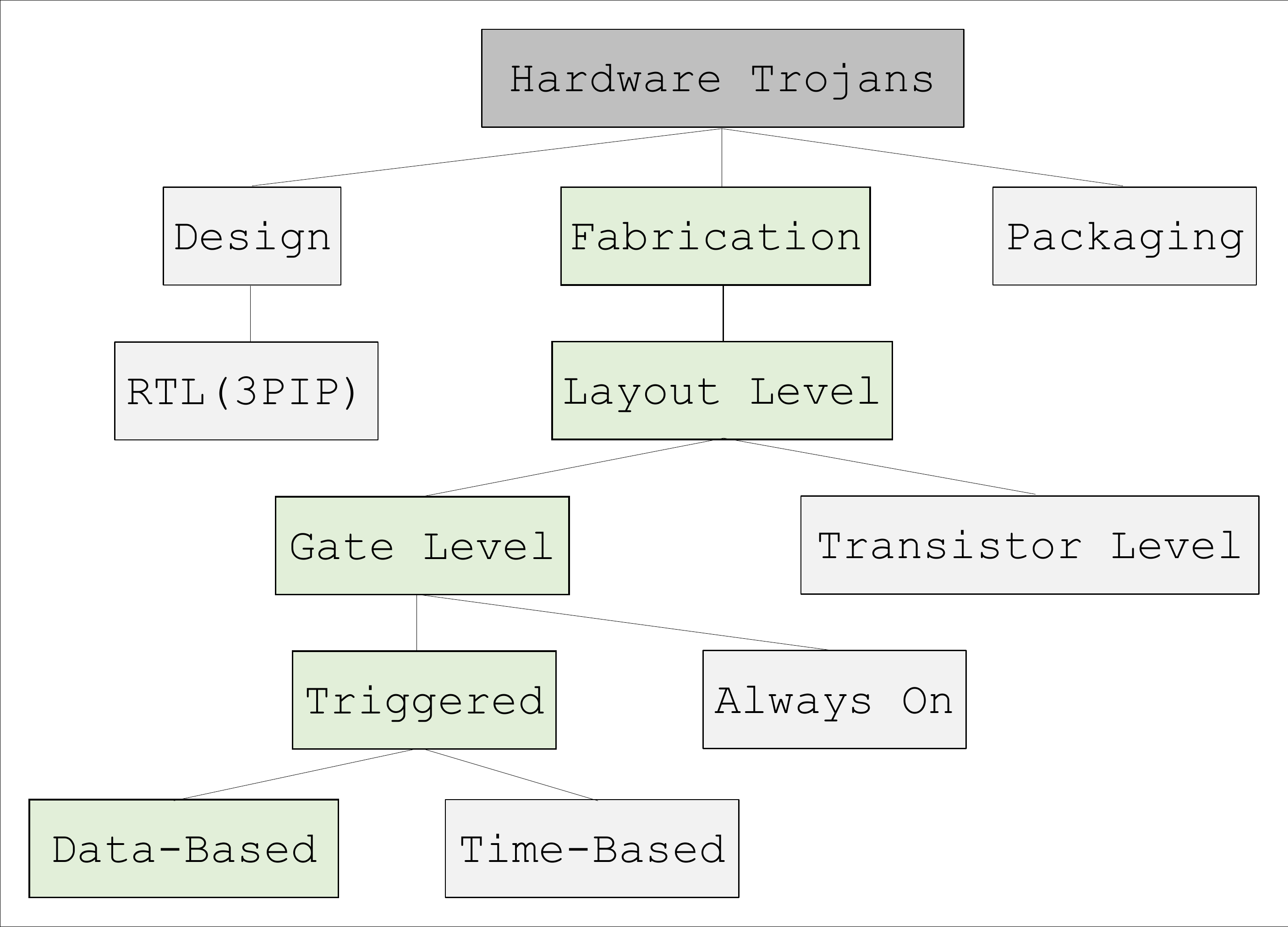}
\caption{Hardware Trojan taxonomy~\cite{xiao2016Trojan_survey,swarup_proceedings,trippel2021bomberman}.}
\label{fig:taxonomy}
\end{figure}

\begin{figure}[tb]
\centering
\includegraphics[trim=0.2cm 0.2cm 0.2cm 0.6cm, clip, width=0.475\textwidth]{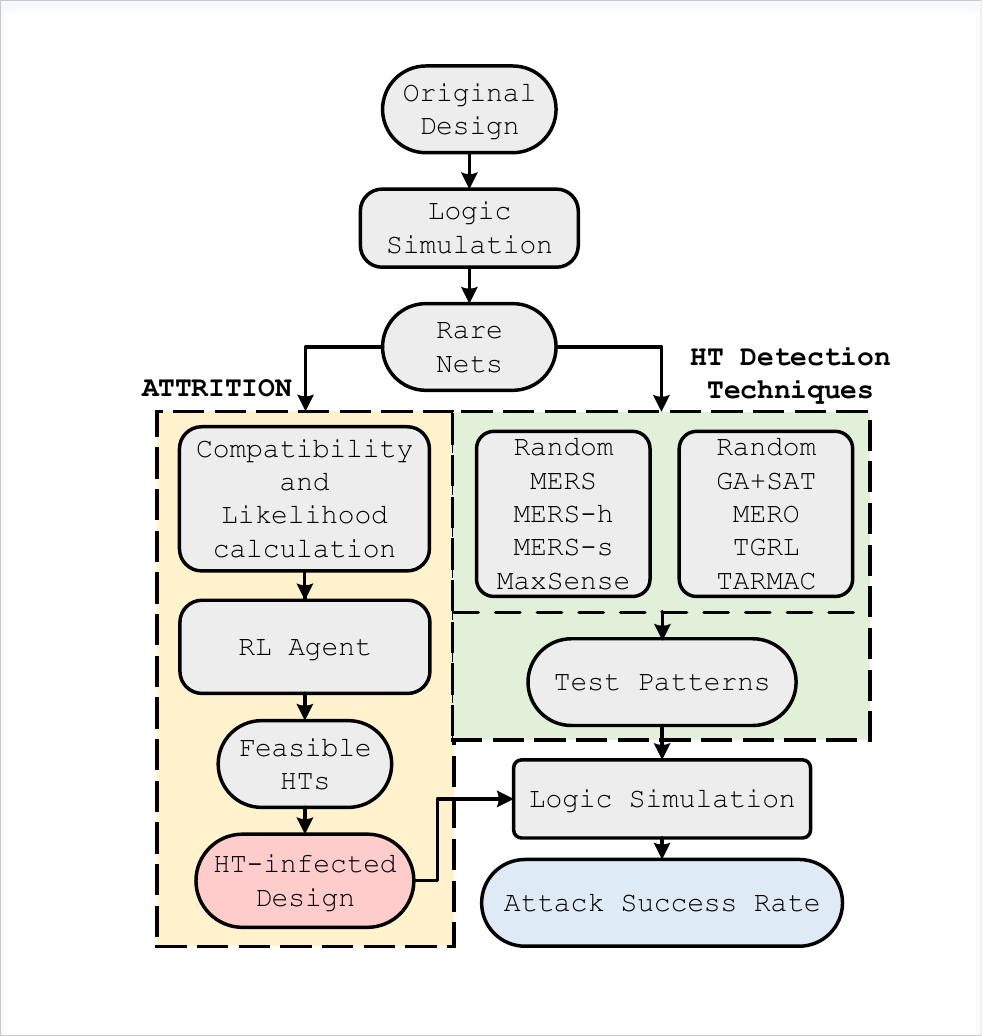}
\smallerspacecaption
\smallerspacecaption
\smallerspacecaption
\smallerspacecaption
\smallerspacecaption
\caption{A high-level overview of the experimental flow.}
\label{fig:exp_flow}
\end{figure} 

\subsection{Hardware Trojan Taxonomy}
\label{app:appendix_HT_taxonomy}

To better understand hardware Trojans (HTs) and generate effective defense techniques, we need to understand the taxonomy of HTs~\cite{swarup_proceedings,xiao2016Trojan_survey} (Figure~\ref{fig:taxonomy}). 
We classify HTs based on three attributes: (i)~insertion phase, (ii)~abstraction level, and (iii)~activation mechanism.
The insertion phase denotes where an HT can be inserted by an adversary in the IC supply chain, including design, fabrication, and packaging.
Since we consider fabrication-time HTs, \textit{i.e.,} HTs inserted by an adversary during the fabrication of ICs, two options emerge.
Either an adversary can insert an HT by adding extra logic gates or performing stealthy manipulations on the transistors.
In our work, we consider gate-level HTs consistent with most prior works in HT research.
Finally, we consider the activation mechanism of HTs, \textit{i.e.,} how an HT will be activated in-field.
The triggering mechanisms include always-on and triggered.
Always-on HTs have constant-time effects on the chip operation as long as the power is supplied. 
Triggered HTs are activated through input patterns. 
External stimulation can be classified into time-based and data-based. 
Time-based stimulus usually is controlled and applied by counters.
Data-based stimulus refers to a single or a sequence of input data patterns.

\subsection{Experimental Flow}
\label{app:appendix_experimental_flow}

Figure~\ref{fig:exp_flow} provides a high-level overview regarding our experimental flow.
First, we start with an HT-free design. 
We perform logic simulations on the design to obtain the probabilities of each wire in the design being $0$ or $1$. 
Then, according to a pre-determined \textbf{rareness threshold}, we obtain the \textbf{rare nets} (\textit{i.e.,} nets whose probability of being $0$ or $1$ is lower than the rareness threshold) from the design. 
Finally, our RL agent uses these rare nets to generate stealthy HTs.
To do so, we first perform the compatibility and likelihood calculations.
Then, we train the RL agent, and the HTs are generated once the agent training ends. 
These HTs are then inserted into the HT-free design to generate HT-infected designs. 
On the other hand, the rare nets are used by both categories (logic testing- and side channel-based) of detection techniques. 
Next, we generate test patterns from the detection techniques to detect HTs. 
Finally, we simulate the HT-infected designs with these test patterns to evaluate the success rate of our attack.

\subsection{Resilience of~\mytool~generated HTs}
\label{app:appendix_resilience_RL_HTs}

\begin{table}[tb]
\centering
\caption{Hyperparameters for~\mytool.}
\label{tab:para}
\resizebox{0.48\textwidth}{!}{
\begin{tabular}{cc}
\toprule
\multicolumn{1}{c}{Parameter}   &  Value\\ 
\midrule
\midrule
\multicolumn{1}{c}{Learning Rate}   & 
    $\begin{dcases}
        0.0003, & \text{if } \left(\text{\# gates} < 100K\right)\\
        0.003, & \text{if } \left(\text{\# gates} \geq 100K\right) \land \left(\text{\# timesteps} <10K\right)\\
        0.0003, & \text{if } \left(\text{\# gates} \geq 100K\right) \land \left(\text{\# timesteps} \geq10K\right)
    \end{dcases}$ \\ 
    \midrule
\multicolumn{1}{c}{Optimizer} &              Adam         \\ \midrule
\multicolumn{1}{c}{Max. Num. of Timesteps} &              $100K$         \\ \midrule
\multicolumn{1}{c}{Num. of Epochs} &               $10$      \\ \midrule
\multicolumn{1}{c}{Num. of Steps} &   $50$                   \\ \midrule
\multicolumn{1}{c}{Batch Size} &          $64$             \\ \midrule
\multicolumn{1}{c}{Discount Factor} &          $0.99$       \\ \midrule
\multicolumn{1}{c}{Clipping Value} &          $0.2$          \\ \bottomrule
\end{tabular}
}
\end{table}
\begin{figure*}[b]
    \centering
    \includegraphics[trim=0.1cm 0.1cm 0.1cm 0.1cm, clip, width=\textwidth]{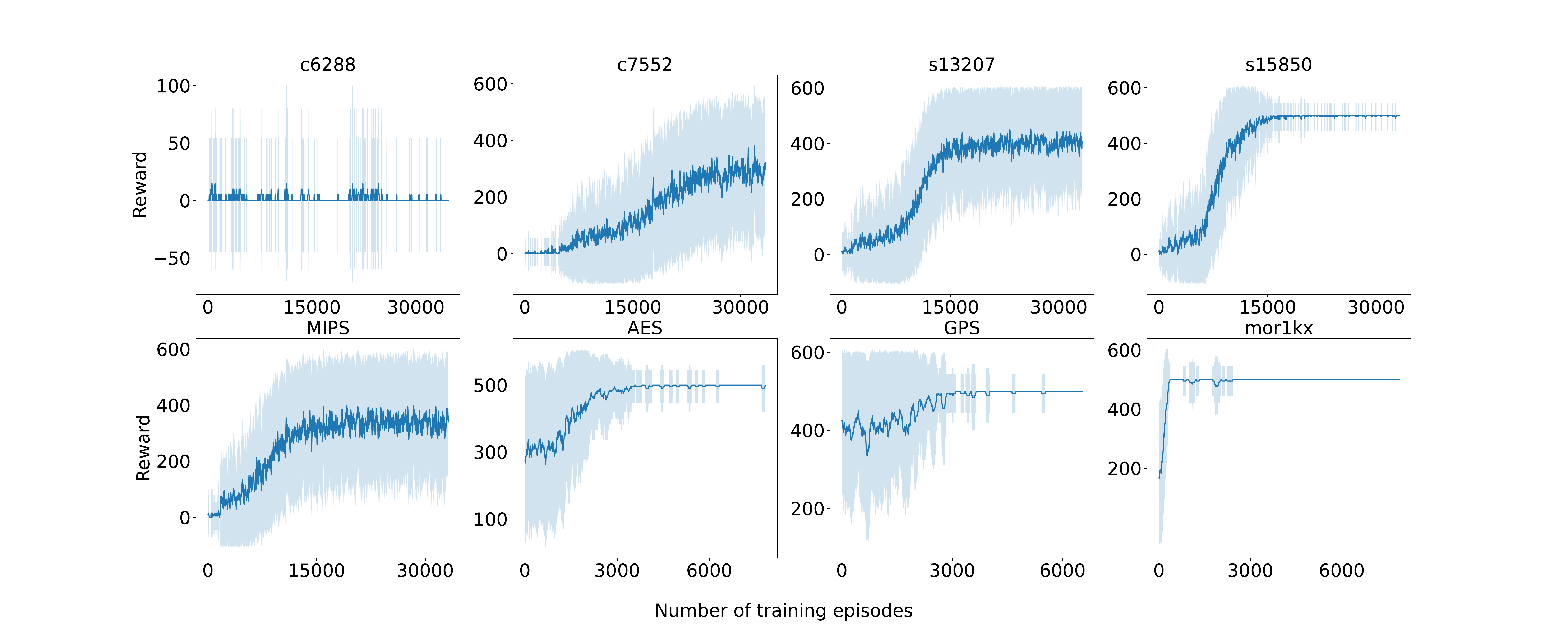}
    \smallerspacecaption
    \smallerspacecaption
    \smallerspacecaption
    \caption{Rewards vs. number of training episodes. Shaded region represents the standard deviation of reward.}
    \label{fig:rew_plot_all}
\end{figure*}

Designers can potentially defend chips against fabrication-time HTs using (i)~integrated circuit (IC) fingerprinting~\cite{agrawal2007trojan} and (ii)~on-chip sensors~\cite{li2008speed,forte2013temperature}.

IC ﬁngerprinting uses side-channel information (e.g., path delay, switching power, etc.) to model the behavior of a golden (\textit{i.e.,} HT-free) IC.
Input patterns are applied to HT-free and HT-infested ICs, and measurements are collected.
If the difference between the collected measurements exceeds a certain threshold (as determined by the defender), the chip is labeled untrustworthy.
Researchers in~\cite{agrawal2007trojan} state that their approach requires the HTs to be at least 0.01\% of the circuit---\mytool~generated HTs occupy (a maximum) of 0.006\% of the overall chip area, thus, they cannot be detected.

Note that fingerprinting-based techniques suffer from a significant limitation, \textit{i.e.,} the availability and reliance on a golden, trusted IC. 
As a result, researchers developed approaches that replace the golden chip with self-referencing~\cite{narasimhan2011tesr}.
For example, temporal self-referencing~\cite{narasimhan2011tesr} compares the transient current of an IC with itself across different time intervals. 
Unfortunately, the ability of temporal self-referencing (in HT detection) is contingent on finding input patterns that activate the HT.
MERS~\cite{huang2016mers} is one of the techniques for generating such input patterns. However, as our results show,~\mytool~generated HTs have minimal impact on switching; hence, they are resistant to self referencing-based methods.

Adding on-chip sensors such as temperature/thermal sensors can also
detect fabrication-time HTs~\cite{forte2013temperature}. 
These sensors detect deviations in power/thermal profiles caused by HT activation. 
Recall that our~\mytool~generated HTs require a minimum of $6$ to $10$ logic gates for a circuit with $> 150,000$ gates.

\subsection{Hyperparameters and Reward Trends}
\label{app:additional_results_1}

For an interested reader, we demonstrate additional information here. Table~\ref{tab:para} lists the hyperparameters of our agent. Figure~\ref{fig:rew_plot_all} shows the reward trends for all benchmarks. 
We observe that for all benchmarks except \texttt{c6288}, the reward curves ramp up to the maximum value quickly, indicating that the agent has learned to generate stealthy HTs. 
The reason for the different reward curve for \texttt{c6288} is due to the structure of the design. \texttt{c6288}, a $16\times16$ multiplier, has several rare nets compatible with each other. 
Hence, it is quite easy for a test pattern to activate several rare nets simultaneously, leading to the agent's inability to find stealthy HTs.

\subsection{Additional Results for Side channel-based Detection Techniques}
\label{app:additional_results_2}

Table~\ref{tab:other_sc_techniques_detection_rate} shows the percentage of HTs that evade random, MERS, MERS-h, and MERS-s detection techniques, \textit{i.e.,} the percentage of HTs with side-channel sensitivity less than 10\%. 
A majority of HTs (both randomly generated and~\mytool~generated) evade all of the above techniques. 

\subsection{Additional Results for Efficacy of Related Work on Inserting Hardware Trojans using Reinforcement Learning}
\label{app:additional_results_for_glsvlsi}

Table~\ref{tab:GLSVLSI_results_trig_wid_breakdown_appendix} shows the attack success rates (against TARMAC) of HTs of different trigger widths from~\cite{sarihi2022hardware}. The results indicate that for all benchmarks, all HTs with trigger width less than 5 are detected by TARMAC, and only a tiny fraction of HTs ($0.15\%$) with trigger width 5 evade detection.

\begin{table}[tb]
\centering
\caption{Number of steps, episodes, and run time of training.}
\label{tab:training_times}
\resizebox{0.46\textwidth}{!}{
\begin{tabular}{ccccccccc}
\toprule
\rotatebox{45}{Design} & \rotatebox{45}{\texttt{c6288}} & \rotatebox{45}{\texttt{c7552}} & \rotatebox{45}{\texttt{s13207}} & \rotatebox{45}{\texttt{s15850}} & \rotatebox{45}{\texttt{MIPS}} & \rotatebox{45}{\texttt{AES}} & \rotatebox{45}{\texttt{GPS}} & \rotatebox{45}{\texttt{mor1kx}} \\
\midrule
\begin{tabular}[c]{@{}c@{}}Num. of\\ Steps\end{tabular} & 100K & 100K & 100K & 100K & 100K & 23.7K & 19.9K & 23.9K \\
\midrule
\begin{tabular}[c]{@{}c@{}}Num. of\\ Episodes\end{tabular} & 35.2K & 33.6K & 33.3K & 33.3K & 33.3K & 7.9K & 6.6K & 7.9K \\
\midrule
\begin{tabular}[c]{@{}c@{}}Training\\ time (hrs.)\end{tabular} & 0.65 & 0.91 & 0.57 & 0.75 & 7.09 & 12 & 12 & 12 \\
\bottomrule
\end{tabular}
}
\end{table}

\begin{table}[tb]
\centering
\caption{Breakdown of attack success rate for HTs of different trigger widths generated by~\cite{sarihi2022hardware} against TARMAC~\cite{TARMAC_TCAD}.}
\label{tab:GLSVLSI_results_trig_wid_breakdown_appendix}
\resizebox{0.46\textwidth}{!}{
\begin{tabular}{cccccc}
\toprule
\multirow{2}{*}{Design} & \multicolumn{5}{c}{Attack Success Rate (\%)} \\
\cmidrule(lr){2-6}
 & 2-width HTs & 3-width HTs & 4-width HTs & 5-width HTs & All HTs\\
\midrule
\texttt{c432} & 0 & 0 & 0 & 0& 0\\
\texttt{c880} & 0 & 0 & 0 & 0& 0\\
\texttt{c1355} & 0 & 0 & 0 & 0.1& 0.02\\
\texttt{c1908} & 0 & 0 &0 &0.6 &0.15 \\
\texttt{c3540} & 0 & 0 & 0 &0 &0 \\
\texttt{c6288} & 0 & 0 & 0 & 0.25& 0.02\\
\midrule
\texttt{Average} & 0 & 0 & 0 & 0.15 & 0.03\\
\bottomrule
\end{tabular}
}
\end{table}

\subsection{Analysis of~\mytool~Training Time}
\label{app:training_time}

Here, we analyze the practicality of using~\mytool~in generating stealthy HTs.
Table~\ref{tab:training_times} depicts the number of steps, episodes, and the training time for each design. 
The solutions we developed (\S\ref{sec:methodology}) enable~\mytool~to perform several thousand episodes ($>\!23K$ on average) through which stealthy HTs are generated in less than $12$ hours.
In fact, for small designs like \texttt{c6288}, \texttt{c7552}, \texttt{s13207}, and \texttt{s15850}, the training ends in less than an hour. 
Medium-sized (\texttt{MIPS}) and large designs (\texttt{AES}, \texttt{GPS}, \texttt{mor1kx}) require about $7$ to $12$ hours of training to generate stealthy HTs.
Even for the largest design (\texttt{GPS} with $>190,000$ gates), the training time is bound by $12$ hours.
This training time is minimal compared to the three months allotted to foundries for fabrication~\cite{ICAS}; hence, an adversary in the foundry has ample time to perform the attack.
Finally, note that our developed solutions (\S\ref{sec:methodology}) enable~\mytool~to perform several thousand episodes ($>23$K on average) through which stealthy HTs are generated in less than $12$ hours.

\begin{table*}[tb]
\centering
\caption{Comparison of percentage of HTs with side-channel sensitivity less than 10\%, for randomly generated and~\mytool~generated HTs against random, MERS, MERS-h, and MERS-s~\cite{huang2016mers}.}
\label{tab:other_sc_techniques_detection_rate}
\resizebox{0.8\textwidth}{!}{
\begin{tabular}{cccccccccc}
\toprule
\backslashbox{Technique}{Design}
& & c6288 & c7552 & s13207 & s15850 & MIPS & AES & GPS & mor1kx \\
\midrule
\multirow{2}{*}{Random} & Random HTs & 100 & 100 & 100 & 100 & 100 & 100 & 100 & 100 \\
 & RL HTs & 100 & 100 & 100 & 100 & 100 & 100 & 100 & 100 \\
\midrule
\multirow{2}{*}{MERS~\cite{huang2016mers}} & Random HTs & 100 & 100 & 100 & 100 & 100 & 100 & 100 & 100 \\
 & RL HTs & 100 & 100 & 100 & 100 & 100 & 100 & 100 & 100 \\
\midrule
\multirow{2}{*}{MERS-h~\cite{huang2016mers}} & Random HTs & 83 & 100 & 100 & 100 & 100 & 100 & 100 & 100 \\
 & RL HTs & 96 & 99 & 100 & 100 & 100 & 100 & 100 & 100 \\
\midrule
\multirow{2}{*}{MERS-s~\cite{huang2016mers}} & Random HTs & 80 & 100 & 100 & 100 & 100 & 100 & 100 & 100 \\
 & RL HTs & 96 & 100 & 100 & 100 & 100 & 100 & 100 & 100\\
 \bottomrule
\end{tabular}
}
\end{table*}

\begin{figure*}[tb]
\centering
\includegraphics[width=\textwidth]{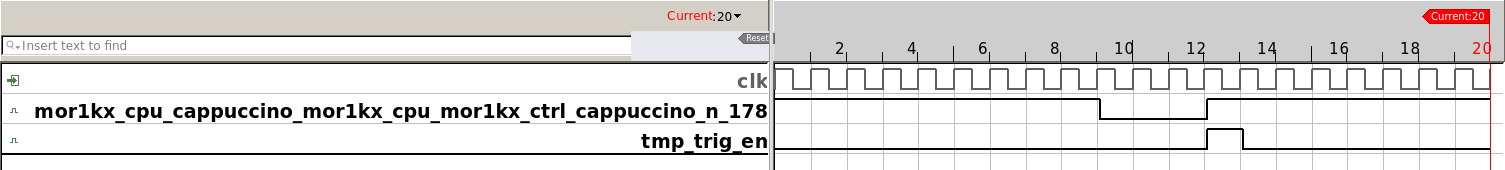}
\smallerspacecaption
\smallerspacecaption
\smallerspacecaption
\caption{Simulation waveform for performing privilege escalation in \texttt{mor1kx} processor.}
\label{fig:mor1kx_priv_esc}
\end{figure*}

\begin{figure*}[tb]
\centering
\includegraphics[width=\textwidth]{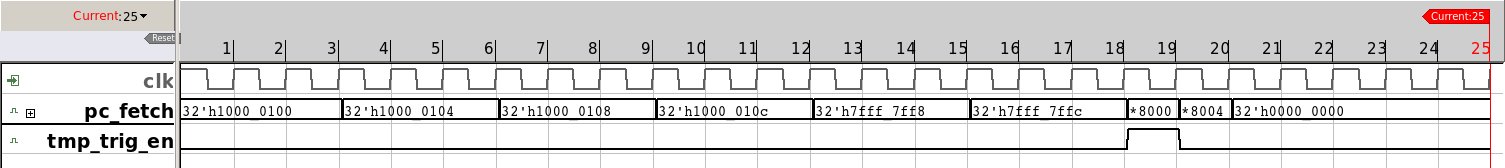}
\smallerspacecaption
\smallerspacecaption
\smallerspacecaption
\caption{Simulation waveform for performing kill switch operation in \texttt{mor1kx} processor.}
\label{fig:mor1kx_kill_switch}
\end{figure*}

\subsection{Waveforms for Case Studies}
\label{app:additional_results_3}

Figure~\ref{fig:mor1kx_priv_esc} shows the simulation waveform for the privilege escalation attack on \texttt{mor1kx}. The \texttt{mor1kx\_cpu\_cappuccino\_mor1kx\_cpu}\\\texttt{\_mor1kx\_ctrl\_cappuccino\_n\_178} signal is the privilege bit of the processor, and the \texttt{tmp\_trig\_en} signal is the trigger signal of the HT.
Initially, the processor starts in the supervisor mode (\textit{i.e.,} privilege bit is $1$) because the processor is designed in that way. 
Hence, to demonstrate a privilege escalation, we make the privilege bit $0$ and then activate the HT. 
The HT is activated in the 13\textsuperscript{th} clock cycle, and immediately, the privilege bit escalates to $1$, indicating a successful attack.

Similarly, Figure~\ref{fig:mor1kx_kill_switch} shows the simulation waveform for the kill switch attack that prevents the processor from executing any useful instructions. The \texttt{pc\_fetch} signal shows the $32-$bit program counter, and the \texttt{tmp\_trig\_en} signal is the trigger that activates the HT. 
The HT is activated in the $19^{\text{th}}$ clock cycle, and impact of the HT is propagated to the target register, \textit{i.e.,} the program counter, in the $21^{\text{st}}$ clock cycle. 
After that, the program counter gets stuck at \texttt{0x00000000}, preventing further execution.

\subsection{Reinforcement Learning for Attacks}
\label{app:RL_attacks}

Here, we discuss how RL has been used by researchers to develop attacks in the general domain of security.
\cite{zhao2021structural} proposed the first structural attack against graph-based Android malware detection techniques.
~\cite{song2020mab} formulated a RL-guided framework that performs adversarial attacks in a black-box setting on state-of-the-art machine learning models for malware classification.
~\cite{zhang2022semantics} proposed a semantics-preserving RL-based attack against black-box graph neural networks for malware detection, achieving a significantly higher evasion rate than four state-of-the-art attacks.
~\cite{li2022reinforcement} showcased a RL-enabled malicious energy attack whereby judiciously choosing intelligent nodes in a green IoT network, an adversary can encourage the data traffic passing through a compromised node, thereby harming information security.

To the best of our knowledge, developing attacks using RL for hardware security problems has received little to no attention.
In that light,~\mytool~showcases flawed security assumptions made by prior and state-of-the-art HT detection techniques across logic testing and side channel approaches by inserting stealthy HTs.

\end{document}